\documentclass[%
reprint,
 amsmath,amssymb,
 aps,
]{revtex4-2}

\usepackage{graphicx}
\usepackage{dcolumn}
\usepackage{bm}
\usepackage{enumitem} 
\usepackage{svg}
\usepackage{subcaption}

\begin{document}


\title{Coexistence vs collapse in transposon populations}

\author{Aria Yom}
\affiliation{
    Department of Physics, University of California, San Diego.
}

\author{Nathan E. Lewis}
\email{natelewis@uga.edu}
\affiliation{
    Departments of Pediatrics and Bioengineering, University of California, San Diego. \\
    Center for Molecular Medicine, Complex Carbohydrate Research Center, and Department of Biochemistry and Molecular Biology University of Georgia. 
}

\date{\today}

\begin{abstract}
Transposons are small, self-replicating DNA sequences found in every branch of life. Often, one transposon will parasitize another, forming a tiny intracellular ecosystem. In some species these ecosystems thrive, while in others they go extinct, yet little is known about when or why this occurs. Here, we present a stochastic model for these ecosystems and discover a transition from stable coexistence to population collapse when the propensity for a transposon to replicate comes to exceed that of its parasites. Our model also predicts that replication rates should be low in equilibrium, which appears to be true of many transposons in nature.
\end{abstract}

\maketitle

\section{Introduction}

Transposons are tiny intragenomic parasites found in every branch of life \cite{Siguier_14_Chandler, Thomas_15_Pritham, Wells_20_Feschotte}. In humans, they comprise over 40\% of our DNA and can contribute to a number of diseases, such as hemophilia and cancer \cite{Lander_01, Kazazian_88_Antonarakis, Pradhan_22_Ramakrishna}. Unlike most pathogens, transposons are rarely exchanged between hosts, passing instead from host to offspring like any typical gene. The uniqueness, ubiquity, and simplicity of the transposon makes it an ideal candidate for biophysical modeling.

Two major fissures quarter the landscape of transposon species: the Class I / Class II divide and the distinction between autonomous and non-autonomous elements. Class I elements replicate through an RNA template, while Class II elements replicate entirely as DNA. Autonomous elements encode all the proteins necessary for their replication, while non-autonomous elements must steal these resources from their autonomous counterparts in order to reproduce. It is common for transposons of both classes to come in autonomous/non-autonomous pairs \cite{Prak_00_Kazazian, Havecker_04_Voytas, Venkatesh_20_Nandini}.

The non-autonomous element parasitizes the autonomous element by consuming its reproductive machinery. The autonomous element in turn parasitizes the host cell. Researchers have thus advocated an ``ecological'' view of elements cohabiting the same host species \cite{Leonardo_02_Nuzhdin, Brookfield_05}. In particular, it has been suggested that the autonomous/non-autonomous element interaction could stabilize transposon populations, much like the predator/prey interaction stabilizes populations in the Lotka-Volterra model \cite{Leonardo_02_Nuzhdin}.

Unfortunately, early models found that when autonomous and non-autonomous elements cohabited a host species, one or both of them went extinct \cite{Kaplan_85_Langley, Brookfield_91, Brookfield_96}. Later models for Class II ecosystems found stable coexistence only with the introduction of an ad hoc nonlinear fitness function to prevent populations from diverging \cite{LeRouzic_06_Capy, LeRouzic_07_Capy}, a precondition unlikely to play a role in stabilizing most natural populations (see Appendix \ref{losses}). Of note, however, Ref. \cite{Xue_16_Goldenfeld} found stable populations in a simple linear model for Class I element ecosystems. Thus, past models disagree over the fate of transposon ecosystems.

In nature, some transposons persist stably over long timescales, while others flare up sporadically or go extinct \cite{Feschotte_02_Wessler, Oggenfuss_21_Croll}. These two behaviors mirror the two dynamics described in past models. One may therefore wonder, what exactly determines the stability or instability of transposon ecosystems?

Here, we probe this question by way of a general model for interacting transposon strains in sexually reproducing hosts. We derive our model from first principles based on the life cycle of a simple Class II element known as a helitron. In this case, we find that ecosystems achieve a stable equilibrium only when the non-autonomous elements' ability to obtain reproductive resources exceeds that of the autonomous elements. This criterion divides transposon populations into two phases, a coexisting phase and a collapsing phase. We also show that transposition rates and autonomous/non-autonomous element ratios should be low in equilibrium, which appears to be true of many transposons in nature.

\section{Model Derivation} \label{model}

\begin{figure}
    \centering
     \includegraphics[width = \columnwidth]{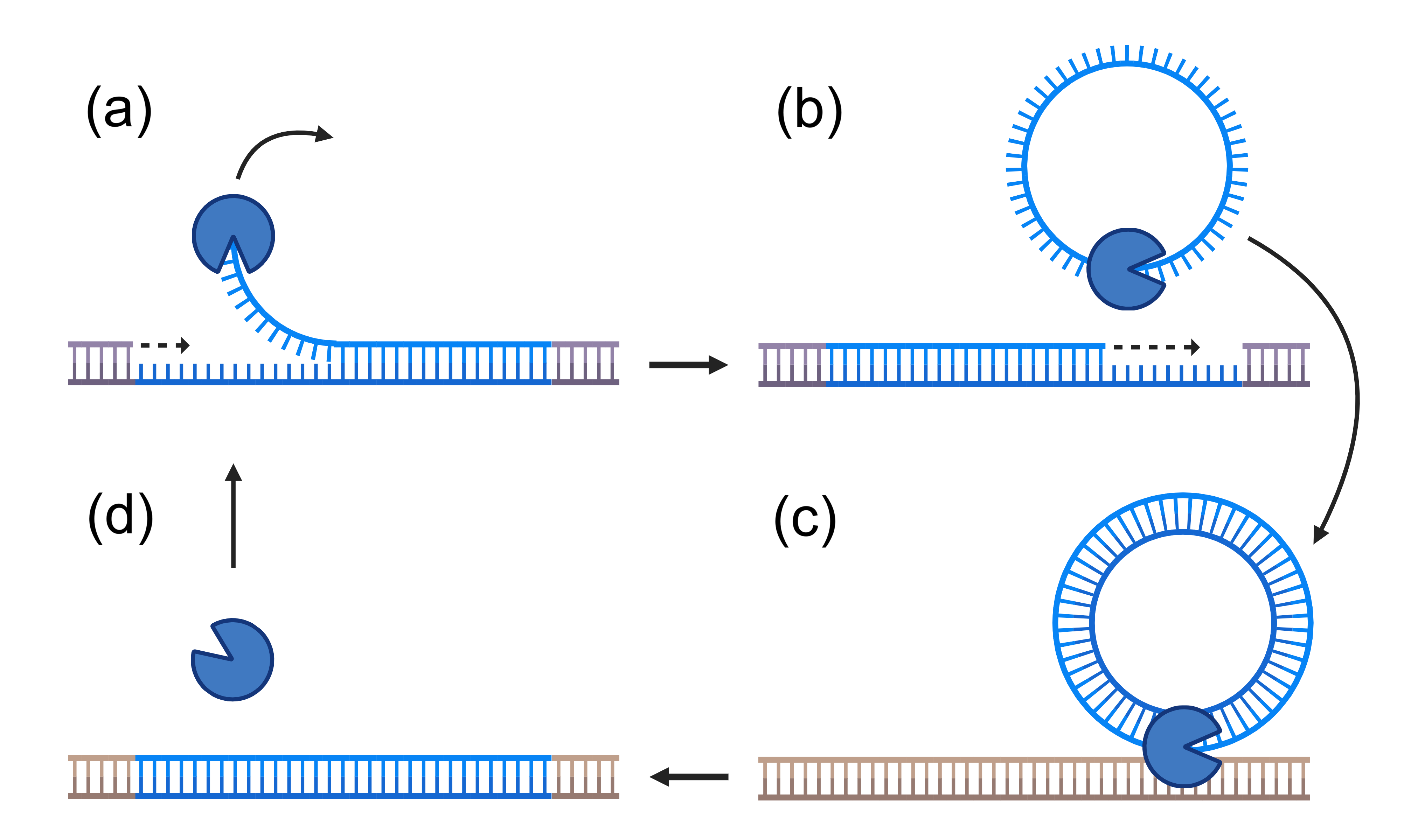}
     \caption{Helitron replication mechanism. (a) The transposase binds to one end of the helitron and begins peeling one strand. (b) The transposase excises and circularizes the single strand of helitron DNA. The cell repairs the missing material. (c) The helitron is converted into double stranded form and moved to a new location in the genome. (d) The helitron is inserted into a new genomic location. Note that some details of this process remain unknown. Please see Refs. \cite{Grabundzija_16_Ivics, Grabundzija_18_Dyda, Chakrabarty_23_Sengupta} for more details.}
     \label{fig:copypasta}
\end{figure}

For concreteness, we base our model on the life cycle of a simple type of transposon known as a helitron, which uses a single transposase protein to replicate itself (Figure \ref{fig:copypasta}, reviewed in Refs. \cite{Thomas_15_Pritham, BarroTrastoy_24_Kohler}). Upon binding to the helitron, the transposase copies the helitron DNA sequence and inserts this copy elsewhere into the genome, potentially harming the host. Autonomous helitrons produce transposase, while non-autonomous helitrons must steal it from their autonomous counterparts. Since this process requires only one transposase, the mean transposition rate should scale linearly with the transposase concentration, at least up to a point.

There are four processes for which any model of transposon populations must account:

\begin{enumerate}
    \item Sex
    \item Transposition
    \item Transpositional toxicity
    \item Transposon loss
\end{enumerate}

In the sections that follow, we derive the effects of each of these four processes on the distribution of transposons in our host species. We will take the host population to be infinite until Section \ref{fin}, wherein we analyze finite size effects. For simplicity, we assume that mating is random and indiscriminate among members of the host population. We also assume that mating occurs much more frequently than the per-element rate of transposition, which appears to be true for many real-life transposons in equilibrium \cite{Shen_87_Kleckner, Harada_90_Mukai, Nuzhdin_94_Mackay, Suh_95_Harada, Maside_00_Charlesworth, Sousa_13_Gordo}. The effects of toxicity and loss are also assumed to be small, as is typically the case in nature \cite{Eanes_88_Ajioka, Charlesworth_89_Langley, Houle_04_Nuzhdin, Pasyukova_04_Mackay}.

We define a \emph{strain} $i$ to be a population of transposons with a defined set of parameters $\alpha_i$, $\omega_i$, and $r_i$ that characterize its propensity to replicate, to produce transposase, and to be lost from the genome, respectively. We analyze ecosystems of many interacting strains, but we do not consider events wherein one strain mutates into another.

We shall conclude this section by deriving an equation for the complete dynamics of transposon populations in our model. Let us therefore begin by computing the effect of each of these subprocesses on the mean transposon count.

\subsection{Sex}

Let $\phi_i$ be a random variable denoting the number of elements of strain $i$ in a randomly chosen host. So long as the population is in linkage equilibrium, which occurs whenever mating is random, frequent, and indiscriminate among hosts, $\phi_i$ will be Poissonian \cite{Brookfield_96}. It follows that we may compute the entire evolution of the transposon distribution simply by accounting for changes in the mean $\lambda_i = \langle \phi_i \rangle$. Random mating has no effect on $\lambda_i$, so let us move on to processes which do.

\subsection{Transposition}

We assume that each strain $i$ in this ecosystem replicates at a rate proportional to $\alpha_i$ and produces transposase at a rate $\omega_i$. The sum $\sum_j \omega_j \phi_j$ quantifies the total amount of transposase being produced within the cell, and acts as a background field stimulating the replication of transposons. For simplicity, and because the half-life of a transposase is typically much smaller than the replicative timescale \cite{Steiniger_06_Reznikoff, Bire_13_RouleuxBonnin, Querques_19_Barabas}, we imagine that the transposase production rate and the cellular transposase concentration are proportional. The probability of replication for a strain $i$ will thus be given by $\frac{\delta p_i}{\delta t} = \alpha_i \phi_i \sum_j \omega_j \phi_j R\left(\sum_k \alpha_k \phi_k \right)$, where we have introduced the functional response $R$ to account for the reduction in transposition that occurs when elements must compete for a limited amount of transposase.

In our model, we take the functional response $R$ to be Holling's type II, $R(x) = \frac{a}{b + c x}$, although the precise form of this function is not important. What matters is that it decays like $\frac{1}{x}$, which should be true of any model, since at some point the replication rate must be bounded by the availability of free transposase or another resource. Without loss of generality, we may absorb the constants $a, b, c$ into $\alpha_i$ and $\omega_i$, to get $R(x) = \frac{1}{1 + x}$. Let us now proceed to calculate the change in the mean transposon count $\lambda_i$:

\begin{align}
&\left( \frac{d \lambda_i}{dt} \right)_\text{trans} = \alpha_i \left\langle \frac{\phi_i \sum_j \omega_j \phi_j}{1 + \sum_k \alpha_k \phi_k} \right\rangle \label{eqn:forst} \\
&= \alpha_i \lambda_i  \left\langle \frac{\omega_i}{1 + \sum_k \alpha_k \phi_k + \alpha_i } +  \sum_j \frac{ \omega_j \lambda_j}{1 + \sum_k \alpha_k \phi_k + \alpha_i + \alpha_j }\right\rangle \label{eqn:22} \\
&= \frac{\alpha_i \lambda_i \left(\omega_i + \sum_j \omega_j \lambda_j\right)}{1 + \sum_k \alpha_k \lambda_k} \left( 1 +  O\left( \frac{\max_l \alpha_l}{\sum_m \alpha_m \lambda_m} \right) \right) \\
&\sim \frac{\alpha_i \lambda_i (\omega_i + \Omega)}{1 + A} \label{eqn:squib}
\end{align}

where we have defined $A = \sum_k \alpha_k \lambda_k$,  $\Omega = \sum_k \omega_k \lambda_k$, and assumed $\alpha_j \ll \sum_k \alpha_k \phi_k \enspace \forall j$, which is true when the total transposon activity eclipses the activity of any single element in the cell. The brackets $\langle  \enspace \rangle$ denote the Poisson expectation value, and we have employed the formula $\langle \phi_i f(\phi_i) \rangle = \lambda_i \langle f(\phi_i + 1) \rangle$ to obtain Equation (\ref{eqn:22}). Note that the approximation (\ref{eqn:squib}) is valid because $\frac{1}{1 + A} \sim \frac{1}{A}$ for large $A$, and because we expect the mean number of helitrons per host to be large \cite{Yang_09_Bennetzen, Yang_09_Bennetzen_2, Du_09_Dooner, Xiong_14_Du}. We shall use this approximation with abandon throughout our analysis.

\subsection{Transpositional toxicity}

In the process of replication, a transposon may kill its host with some probability $q$. The probability of replication for a single strain is $\frac{\delta p_i}{\delta t} = \frac{\alpha_i \phi_i \sum_j \omega_j \phi_j}{\sum_k \alpha_k \phi_k}$; thus, the probability that the host is killed in this scenario is $\frac{\delta p_\text{death}}{\delta t} = q \frac{\sum_i \alpha_i \phi_i \sum_j \omega_j \phi_j}{\sum_k \alpha_k \phi_k}$.

The result of this small change in the host population will be a slight alteration to the $\phi_i$ distribution. Let $N(\vec{\phi})$ denote the number of hosts carrying $\phi_i$ transposons of each strain $i$. The net effect of a change in $N(\vec{\phi})$ on the mean transposon count $\lambda_i$ will be:

\begin{align}
\delta \lambda_i &= \delta \frac{\sum_{\vec{\phi}} \phi_i N(\vec{\phi})}{\sum_{\vec{\phi}}  N(\vec{\phi})} \\
&= \frac{\sum_{\vec{\phi}} \phi_i\delta N}{\sum_{\vec{\phi}} N} - \frac{\sum_{\vec{\phi}} \phi_i N}{\sum_{\vec{\phi}} N} \frac{ \sum_{\vec{\phi}} \delta N}{\sum_{\vec{\phi}} N} \\
&= \left\langle (\phi_i - \lambda_i) \frac{\delta N}{N} \right\rangle
\end{align}

Thus, the change in $\lambda_i$ will be:

\begin{align}
\left( \frac{d \lambda_i}{dt} \right)_\text{tox} &= - q \left\langle \frac{ (\phi_i - \lambda_i ) \sum_j \omega_j \phi_j \sum_k \alpha_k \phi_k} {1 + \sum_l \alpha_l \phi_l} \right\rangle \label{eqn:hard} \\ 
&\sim - \frac{q \lambda_i (\alpha_i \Omega  + \omega_i A) }{1 + A}
\end{align}

One may also wonder whether the helitron places some fitness burden on its host even when it is not replicating. In that case, we would take the toxicity to be proportional to the number of helitrons and get $\frac{d \lambda_i}{dt} \propto \lambda_i$. As we shall see, such a term can simply be absorbed into the term for transposon loss.

\subsection{Transposon loss}

Finally, we consider the effects of genetic drift and transposon excision. Since each element has some rate $r_i$ of decaying or being lost in each generation, the effect on $\lambda_i$ will be:

\begin{equation}
\left( \frac{d \lambda_i}{dt} \right)_\text{loss} = - r_i \left\langle \phi_i \right\rangle = - r_i \lambda_i
\end{equation}

\subsection{Total mean change in transposon count}

Having considered in detail the individual effects of each of these processes, we may now obtain the total mean change in transposon count:

\begin{equation}
\boxed{ \frac{d \log \lambda_i}{dt} = \frac{\alpha_i (\omega_i + \Omega) - q \omega_i A}{1 + A} - r_i } \label{eqn:main_manz}
\end{equation}

where we have replaced $\sim$ with $=$ since this will serve as our model for the rest of this paper. This equation is valid so long as $\alpha_i \ll \sum_j \alpha_j \lambda_j$ and $q$ and $r_i$ are small.

\section{Results}

\subsection{Stable ecosystems consist of one autonomous and one non-autonomous strain}

\begin{figure}
     \centering
     \includegraphics[width = \columnwidth]{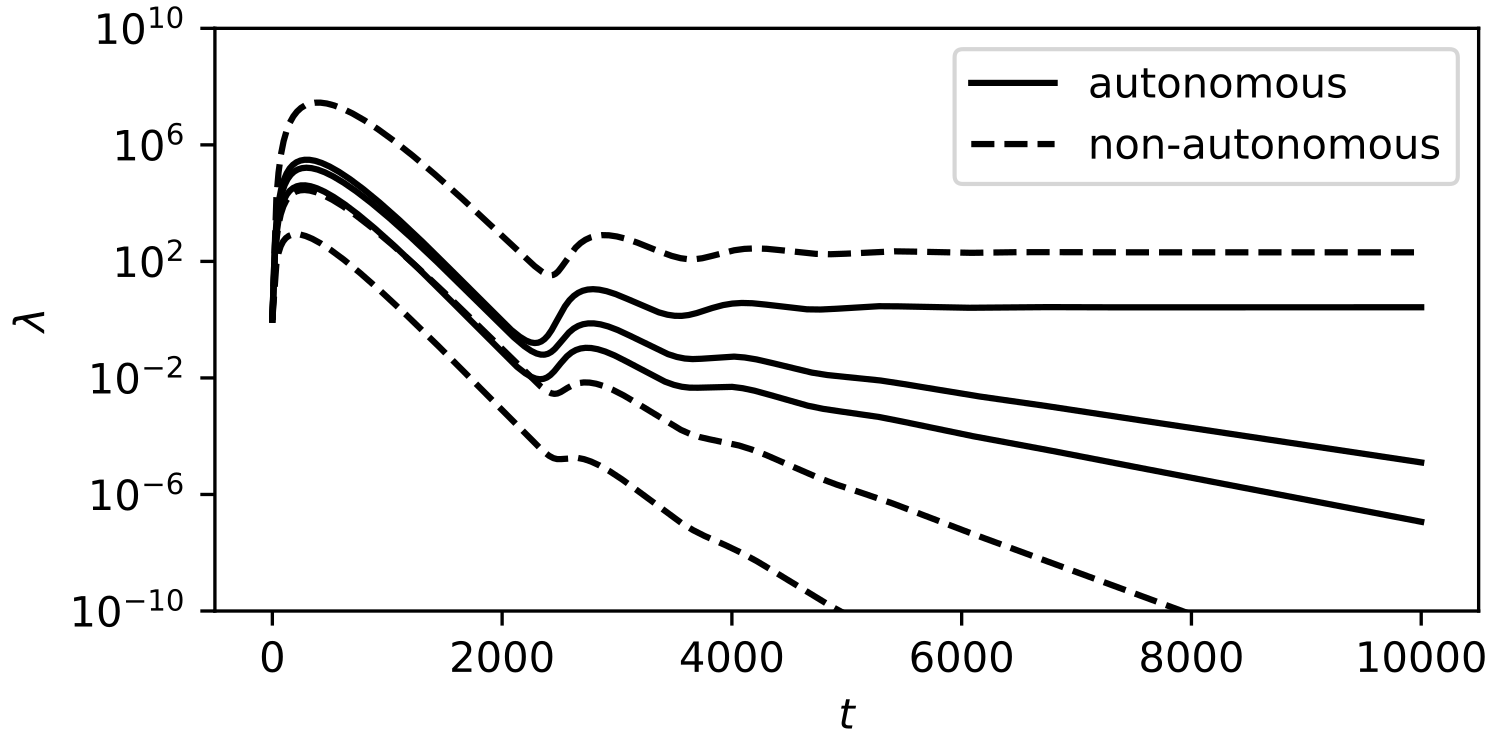}
     \caption{Simulation of a stable system of 3 autonomous and 3 non-autonomous strains, each starting at $\lambda_i(0) = 1$, with $t$ measured in arbitrary units. $\alpha_i$, $\omega_i$, $q$, and $r_i$ were randomly chosen for each strain. In this instance, $\alpha_i = \{ .75, .36, .50, .51, .58, .56\}$, $\omega_i = \{ 0, 0, 0, .61, .77, .42\}$, $q = .00076$, $r_i = .0099$. Note that of the six initial strains, only two persist to $t = \infty$, one autonomous and one non-autonomous.}
     \label{fig:poplotz}
\end{figure}

When do transposon strains coexist in equilibrium in our model? So long as we are not interested in timescales, we may reparameterize our model such that $d t = (1 + A) d \tau$. Equation (\ref{eqn:main_manz}) then simplifies to a generalized Lotka-Volterra model:

\begin{equation}
\frac{d \log \lambda_i}{d \tau} = \alpha_i \omega_i - r_i + \sum_j ( \alpha_i \omega_j - q \omega_i \alpha_j - r_i \alpha_j ) \lambda_j
\end{equation}

If all strains are non-autonomous ($\omega_i = 0$), then there will be no transposase production, and all elements will go extinct. However, if $\omega_i > 0$ for all strains $i$, then since $q$ and $r_i$ are small, $\frac{d \log \lambda_i}{d \tau} > \alpha_i \omega_i$, and $\lambda_i$ will diverge. As $\lambda_i$ grows, the toxicity of the transposons will eventually lead to the collapse of the host population. Therefore, stable populations must consist of at least one autonomous and one non-autonomous strain.

We now proceed to show that, except on a measure-zero subset of our parameter space, stable populations will always consist of precisely one autonomous and one non-autonomous strain. Let us consider the long-time behavior of orbits by defining the long-time averages $\overline{A} = \lim_{T \to \infty} \frac{\int_0^T d \tau A}{T}$ and $\overline{\Omega} = \lim_{T \rightarrow \infty} \frac{\int_0^T d \tau \Omega}{T}$. For any stable strain, $\frac{d \log \lambda_i}{d \tau}$ will average to zero over the long term. Thus:

\begin{equation}
\alpha_i \omega_i - r_i + \alpha_i \overline{\Omega} - (q \omega_i + r_i) \overline{A} = 0
\end{equation}

For these linear equations to yield a solution $(\overline{A}, \overline{\Omega})$, the $3d$ vectors $(\alpha_i \omega_i - r_i, \alpha_i, -q \omega_i -r_i)$ must all lie on the same $2d$ plane. This is not possible without fine-tuning for $n > 2$ strains. It follows that stable ecosystems should consist of exactly one autonomous and one non-autonomous strain. This can be observed in simulations of our model (Figure \ref{fig:poplotz}), wherein less fit strains are whittled away until only one autonomous and one non-autonomous strain remain.

\subsection{Coexistence collapses when $\alpha_a > \alpha_n$}

Having shown that stable orbits consist of one autonomous and one non-autonomous strain, we now restrict our attention to the following two-strain model:

\begin{align}
\frac{d \log \lambda_a}{d \tau} &= \alpha_a - r_a + (1 - q - r_a) \alpha_a \lambda_a - (q + r_a) \alpha_n \lambda_n \\
\frac{d \log \lambda_n}{d \tau} &= - r_n + (\alpha_n - r_n \alpha_n) \lambda_a - r_n \alpha_n \lambda_n
\end{align}
where $a$ and $n$ denote autonomy and non-autonomy, and where we have taken $\omega_a = 1$, which can be done without loss of generality by rescaling $r_a$, $r_n$, $\tau$. Note that this operation makes all of our parameters dimensionless.

The fixed point of these equations occurs at:

\begin{align}
\lambda_a^0 &= \frac{r_n (q + \alpha_a)}{\alpha_n (r_a + q) - \alpha_a r_n} \\
&\sim \frac{r_n \alpha_a}{\alpha_n (r_a + q) - \alpha_a r_n } \\
\lambda_n^0 &= \frac{1}{\alpha_n}\frac{\alpha_a \alpha_n - r_a \alpha_n + r_n \alpha_a (1 - q - \alpha_a) }{\alpha_n (r_a + q) - \alpha_a r_n} \\
&\sim \frac{\alpha_a}{\alpha_n (r_a + q) - \alpha_a r_n}
\end{align}
where we have again employed the approximation $q, r_a, r_n \ll 1$.

Since $\lambda_a, \lambda_n > 0$, this fixed point only exists when $\frac{q + r_a}{\alpha_a} > \frac{r_n}{\alpha_n}$. The existence of a fixed point is our first requirement for coexistence. We must now assess the stability of this fixed point. It is easy to linearize our equations about their fixed point and find the eigenvalues, $\alpha_a - \alpha_n \pm \sqrt{(\alpha_a - \alpha_n)^2 - 4 \frac{\alpha_a \alpha_n^2}{r_n}\left( \frac{q + r_a}{\alpha_a} - \frac{r_n}{\alpha_n} \right)}$. The fixed point is stable when the real parts of these eigenvalues are all negative, which occurs if and only if $\alpha_n > \alpha_a$. To summarize:

$$ \boxed{ \text{Orbits are stable iff} \quad \frac{q + r_a}{\alpha_a} > \frac{r_n}{\alpha_n} \enspace \text{and} \enspace \alpha_n > \alpha_a } $$

\begin{figure}
    \centering
    \includegraphics[width = \columnwidth]{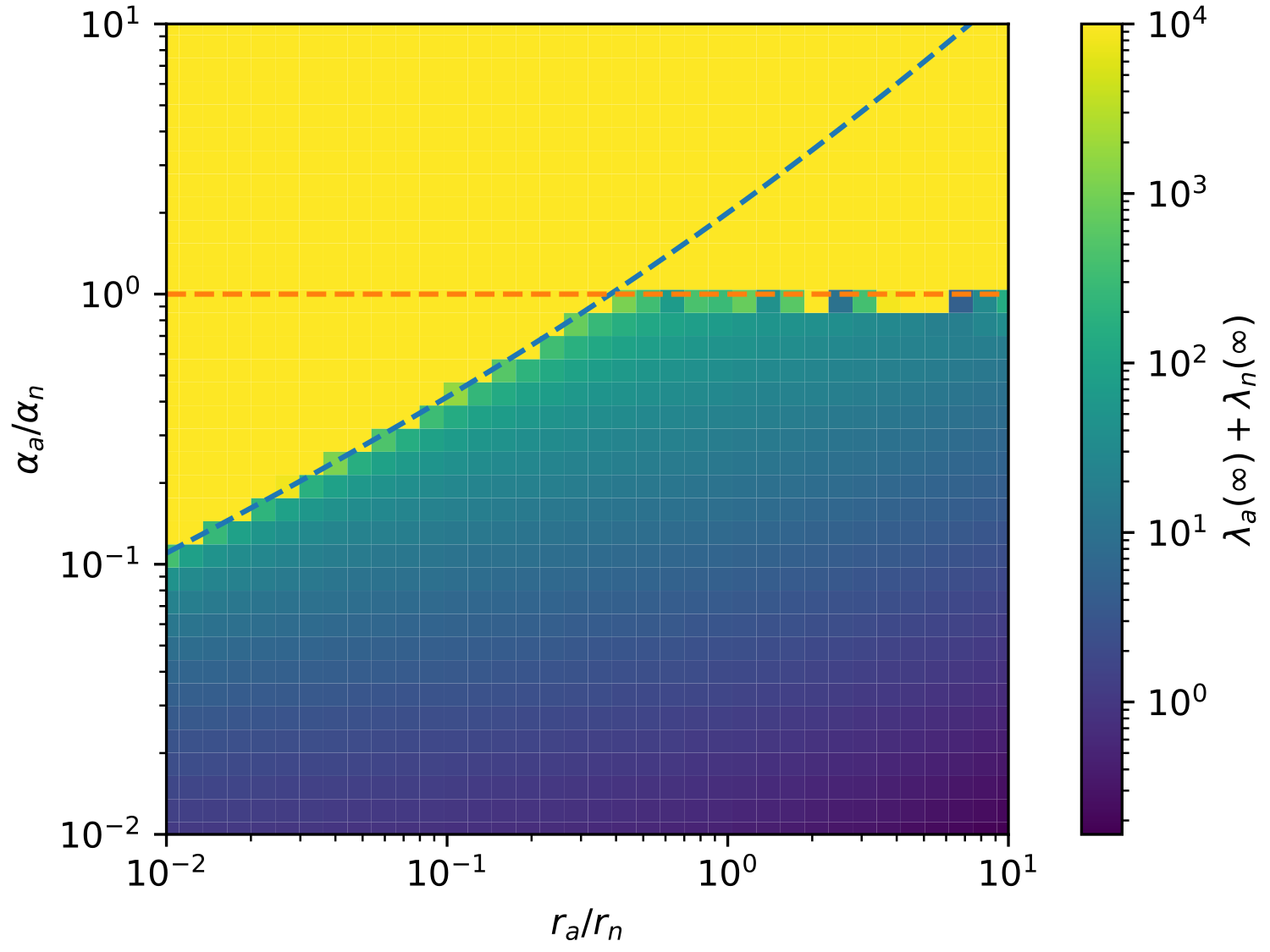}
    \caption{Total transposon counts after long simulations. At each point, $\lambda_a + \lambda_n$ was evaluated at $t=10000$, by which time the system had usually equilibrated. Parameters satisfied $\alpha_a \alpha_n = 1$, $r_a r_n = 10^{-4}$, and $q = 10^{-2}$. The dotted curves denote the phase boundaries $\alpha_a = \alpha_n$ and $\frac{q + r_a}{\alpha_a} - \frac{r_n}{\alpha_n}$.}
    \label{fig:phases}
\end{figure}

We illustrate the transition from stable coexistence to diverging/collapsing orbits in Figure \ref{fig:phases}. It is worth noting that because autonomous elements have more potential points of failure and are more problematic for the cell, we should expect $r_a > r_n$. We may also imagine that the deleteriousness of the transposon is more significant than its decay rate, leading to $q > r_a, r_n$. Either of these conditions will imply that whenever $\alpha_n > \alpha_a$, the model automatically satisfies $\frac{q + r_a}{\alpha_a} > \frac{r_n}{\alpha_n}$. Thus for real life transposons, $\alpha_a = \alpha_n$ is probably the only relevant phase boundary.

\begin{figure*}
     \centering
     \begin{subfigure}{\columnwidth}
         \centering
         \includegraphics[width = \columnwidth]{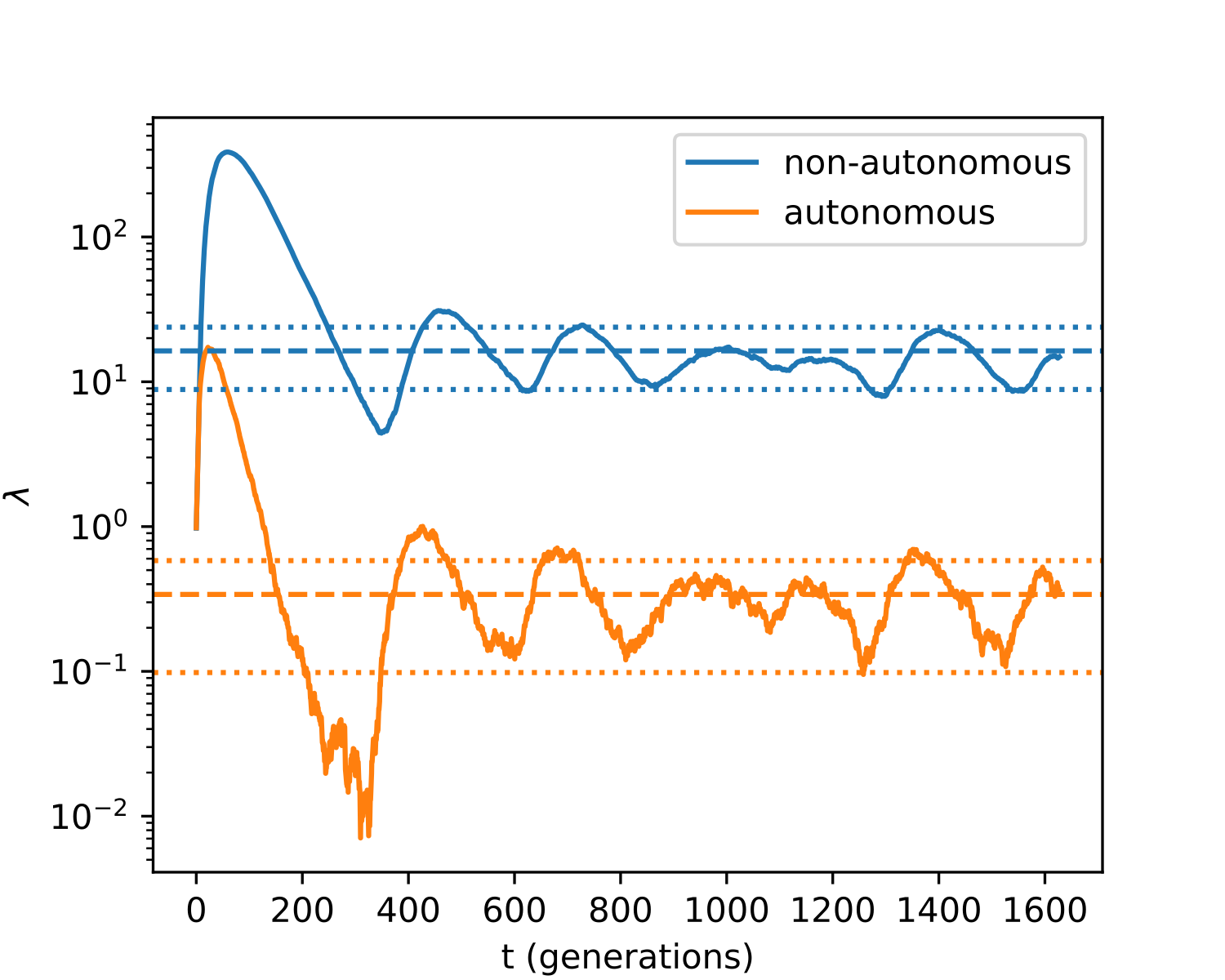}
     \end{subfigure}
     \begin{subfigure}{\columnwidth}
         \centering
         \includegraphics[width = \columnwidth]{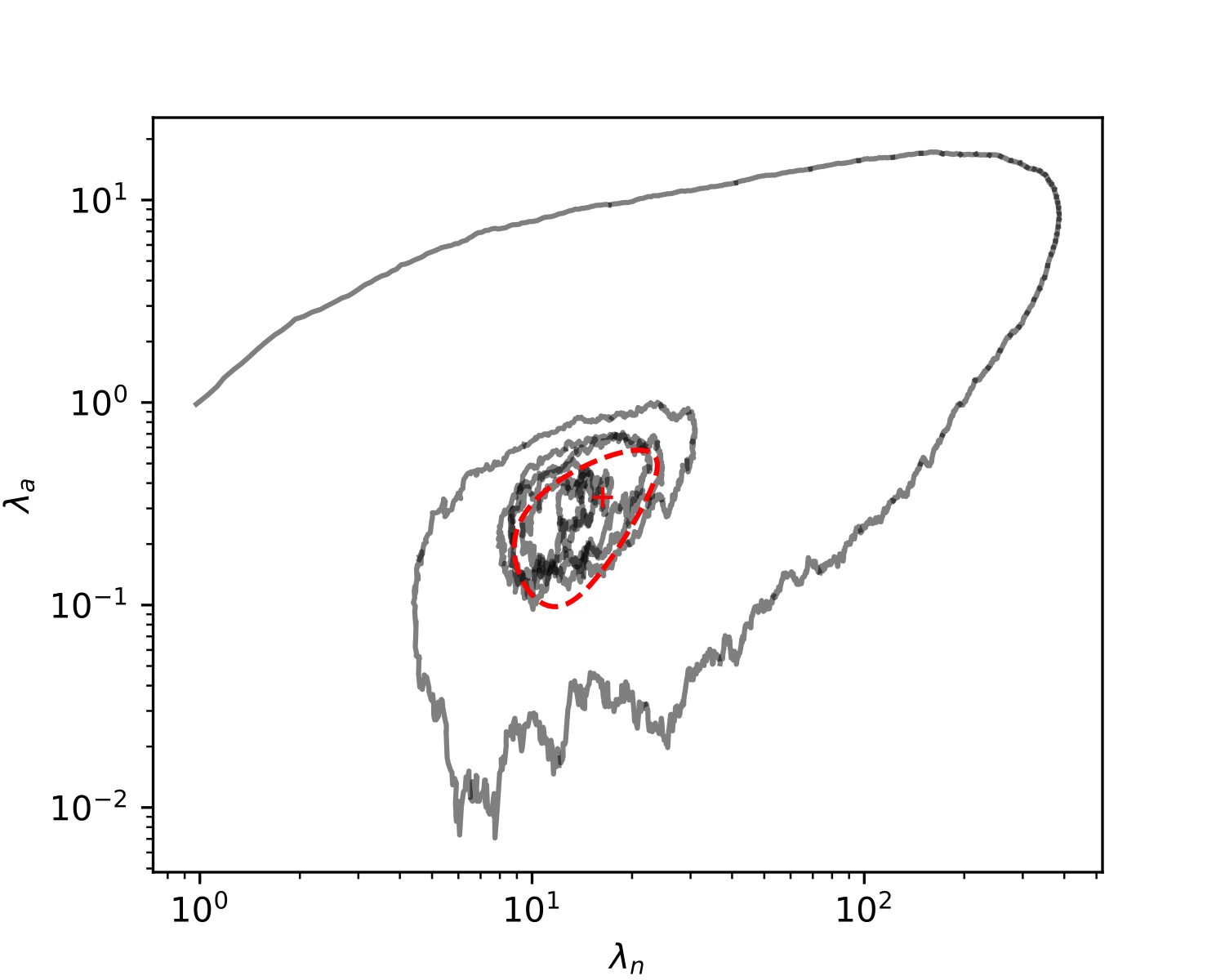}
     \end{subfigure}
     \caption{Simulation of a finite population of autonomous and non-autonomous elements. Parameters used: $\lambda_a(0) = \lambda_n(0) = 1$, $\alpha_a = 1$, $\alpha_n = 2$, $q = r_a = r_n = 10^{-2}$, $N = 4096$. (left) Dynamics of $\lambda_a$ and $\lambda_n$. The infinite population equilibrium values $\lambda_a^0$ and $\lambda_n^0$ are indicated by dashed lines, while the dotted lines denote fluctuations of size $\sqrt{4 \Sigma_{aa}}$ and $\sqrt{4 \Sigma_{nn}}$. (right) Orbit of $\lambda_a$ and $\lambda_n$ with fixed point and $4 \Sigma$ ellipse. }
     \label{fig:bicboi}
\end{figure*}

\subsection{Low transposition rates and few autonomous elements in equilibrium}

Taking the ratio of $\lambda_a$ and $\lambda_n$, we can see that autonomous elements come to be greatly outnumbered by their non-autonomous partners in equilibrium:

\begin{equation}
\frac{\lambda_a^0}{\lambda_n^0} \sim r_n
\end{equation}

As a result, the per-element transposition rate becomes low near the fixed point:

\begin{equation}
\left( \frac{d \log(\lambda_a + \lambda_n)}{dt} \right)_\text{trans} = \frac{\lambda_a}{\lambda_a + \lambda_n} \frac{\alpha_a + A}{1 + A} \sim r_n
\end{equation}

Intriguingly, both of these results that naturally fall out of our model appear to be the norm for helitrons and other transposons in nature \cite{Havecker_04_Voytas, Cordaux_06_Batzer, Bennett_08_Devine, Xing_09_Jorde, Beck_11_Moran, Du_09_Dooner, Yang_09_Bennetzen, Thomas_15_Pritham, Shen_87_Kleckner, Harada_90_Mukai, Nuzhdin_94_Mackay, Suh_95_Harada, Maside_00_Charlesworth, Sousa_13_Gordo}.

\subsection{Finite population effects} \label{fin}

What happens when the number $N$ of hosts is finite? In this case, the randomness of events such as mating cannot be neglected, and the evolution of $\lambda_i$ becomes stochastic. Our system may therefore be described by a multidimensional Fokker-Planck equation whose drift term comes from Equation (\ref{eqn:main_manz}) and whose diffusion term we may compute from the variances of each of the four subprocesses in our model.

Near the fixed point, the only non-negligible contribution to the variance comes from random mating, which contributes a factor of $\sigma^2_i = \frac{2 \lambda_i}{N}$ per generation. As our populations approach equilibrium, the $\lambda_i$ distribution becomes approximately normal, centered about $\lambda_i^0$, with variance $\Sigma_{ij}$. We may solve for $\Sigma_{ij}$ numerically using the Lyapunov equation:

\begin{equation}
J \Sigma + \Sigma J^T + Q = 0
\end{equation}

where $J_{ij} = (\alpha_i \omega_j - (q \omega_i + r_i) \alpha_j ) \frac{\lambda_i^0}{1 + A^0}  $, $Q_{ij} = \frac{2 \gamma \lambda_i^0}{N} \delta_{ij}$, and $\gamma$ denotes the number for generations per unit time, which we may take to be $\gamma = 1$ without loss of generality.

Figure \ref{fig:bicboi} illustrates a sample trajectory from our model. We can see that even for relatively small populations ($N \sim 10^4$), orbits converge to the expected equilibria, and our estimates for the fluctuation sizes are accurate as well.

\section{Discussion}

\subsection{Stability in Class I and instability in Class II models}

We are now in a position to understand why previous Class II models tended towards instability. In each of these models \cite{Kaplan_85_Langley, Brookfield_91, Brookfield_96, LeRouzic_06_Capy, LeRouzic_07_Capy}, the replication rates were taken to be proportional to the transposase concentration, placing them into the same framework analyzed herein. However, in each of these models, the parameters $\alpha_a$ and $\alpha_n$ were taken to be equal for simplicity. Unfortunately, by taking $\alpha_a = \alpha_n$, past researchers placed their populations precisely on the phase boundary between coexistence and collapse. Hence, the instability.

What can we say about the stable equilibrium found in Ref. \cite{Xue_16_Goldenfeld} for long and short interspersed nuclear elements (LINEs and SINEs), a common pair of autonomous and non-autonomous Class I elements? To simplify their model to its mean-field essence, their equations were roughly of the form:

\begin{align}
\frac{d \lambda_a}{dt} &= \alpha_a \Omega - r_a \\
\frac{d \lambda_n}{dt} &= \alpha_n \Omega \lambda_a - r_n 
\end{align}

These equations differ from ours in two important ways: (a) they have no functional response, and (b) the autonomous transposition rate is not proportional to $\lambda_a$. It is the latter difference that accounts for the stability of these equations, since the analogous transposition parameters, $\alpha_a^\text{eff} = \frac{\alpha_a}{\lambda_a}$ and $\alpha_n^\text{eff} = \alpha_n$, do not have a constant ratio. Therefore, if populations begin to diverge, they will eventually reach a regime in which $\alpha_n^\text{eff} > \alpha_a^\text{eff}$, and the system will self-stabilize.

The essential difference between this model and ours arises from the fact that in the LINE replication process, the transposase is imagined to bind immediately to the LINE transcript that generates it, rather than binding to a random element in the DNA \cite{Esnault_00_Heidmann, Wei_01_Moran}. This is known as ``\textit{cis}-preference'' and occurs in some other Class I elements as well, but does not occur in Class II elements \cite{Bolton_05_Boeke, Curcio_15_Lesage}. It follows that \textit{cis}-binding Class I ecosystems should be stable, while Class II ecosystems may be stable or unstable.

\subsection{Predictions of the model}

Aside from the coexistence or collapse of helitron populations, our model also makes the following two quantitative predictions:

\begin{enumerate}[label=(\roman*)]
\item That non-autonomous elements should outnumber autonomous ones, at least in equilibrium.
\item That per-element transposition rates should be low in stable populations.
\end{enumerate}

Our first prediction, although counterintuitive, has been confirmed numerous times \cite{Du_09_Dooner, Yang_09_Bennetzen, Thomas_15_Pritham}. Heuristic arguments for the fitness advantage of non-autonomous elements over autonomous ones have been given by many previous authors (see \cite{Leonardo_02_Nuzhdin} for example). If our results are correct, then we have derived the first formula to quantify this preponderance.

Our second prediction, that per-element transposition rates should be low, has been verified in helitrons and many other transposons \cite{Shen_87_Kleckner, Harada_90_Mukai, Nuzhdin_94_Mackay, Suh_95_Harada, Maside_00_Charlesworth, Sousa_13_Gordo}. One explanation for this phenomenon could be that hosts have evolved to suppress transposons, but if this is the case, then why haven't hosts been able to suppress viruses to a similar extent? The threat posed by viruses is far greater, yet their replication rates remain many orders of magnitude higher. In our model, we show instead that transposons can be kept in check by other transposons rather than by the host.

Do these results hold for transposons other than helitrons? Let us consider the aforementioned case of LINEs and SINEs, which together occupy one third of the human genome \cite{Lander_01}. Since active SINEs outnumber LINEs at least ten to one, they clearly satisfy (i) \cite{Bennett_08_Devine, Beck_11_Moran}. And since both elements transpose on the order of once per hundred generations, amounting some $10^{-4}$ replications per active element per generation, they also satisfy (ii) \cite{Cordaux_06_Batzer, Xing_09_Jorde}. Therefore, these results may be much more universal than one may expect from this simple model.

\subsection{Concluding remarks}

Can this model explain the survival or collapse of real-life transposon populations? This question is difficult to answer, as the relevant parameters for natural populations have not been measured. However, since $\alpha_i$ is essentially a measure of transposase affinity, one could perform an experiment to test our $\alpha_n > \alpha_a$ criterion in any number of transposon species and compare the results to natural populations.

In this article, we derived some simple equations for the stability and properties of transposon ecosystems which appear to explain the behavior of diverse transposon species. In particular, we have shown how subtle parameter changes can cause transposon populations to diverge dramatically, which could shed light on why similar species can have such vastly different genome sizes \cite{Gregory_05}. Transposons are an ancient and ubiquitous part of life on Earth, and are the most abundant elements of our DNA. These results shed light both on the evolution of Eukaryotes and on the dynamics of modern genomes.

All code used to generate our plots is freely available at github.com/moyja/transposon\_paper

\begin{acknowledgments}
This work was supported in part by funding from NIGMS (R35 GM119850). 
\end{acknowledgments}

\appendix

\section{On superlinear toxicities} \label{losses}

Consider a single transposon which replicates at a rate $\beta$ and which causes the host to perish at a rate $\gamma_1$. In this case, the mean rate of change in transposon count will be $ \frac{d \lambda}{dt} = (\beta -\gamma_1) \lambda $. Since such a model admits no nonzero equilibrium, we may be inclined to introduce a nonlinear term to the toxicity, resulting in the following equation:

\begin{equation}
\frac{d \lambda}{dt} = (\beta -\gamma_1) \lambda - \gamma_2 \lambda^2
\end{equation}

This simple approach was advocated in the earliest models for transposon population dynamics to ensure the existence of a stable equilibrium population, and has continued to be used in more recent studies as well \cite{Brookfield_82, Charlesworth_83_Charlesworth, LeRouzic_06_Capy, LeRouzic_07_Capy}. However, in order to obtain such a term, there must be some mechanism by which the transposon toxicity compounds more than linearly in $\lambda$. Such a mechanism can either be transposition-dependent or transposition-independent.

The former case includes effects like insertional mutagenesis. Conceivably perhaps, the transposition of two elements in close succession could kill the host with high probability, while a single transposition would not. But as we mentioned earlier, the most abundant transposons in the human genome replicate just once per thousand years \cite{Cordaux_06_Batzer, Xing_09_Jorde}. It is therefore difficult to imagine how any such interaction could occur.

The latter case includes effects like ectopic recombination, which was once the most popular hypothesis to explain this phenomenon \ \cite{Montgomery_87_Langley, Langley_88_Charlesworth}. In this case it is imagined that the presence of transposons in the genome may disrupt the DNA recombination machinery and lead to fatal errors in crossing over. This view is articulated in Ref. \cite{Brookfield_91}:

\textit{``A balance between transposition and selection can be achieved, but only if the logarithm of the fitness of an individual with $i$ copies of a transposable element declines more than linearly with $i$. This selection could arise through ectopic recombination with the transposable elements acting as scattered sites of homology in the chromosomes. The frequencies of such events will increase with the square of transposable element copy number. One interesting aspect is that the selection arises purely because these sequences are interspersed repetitive, and not because of any transposition process that they are undergoing.''}

There is ample evidence that transposons do indeed cause ectopic recombination, but this only implies a nonzero contribution to $\gamma_1$ \cite{Delprat_09_Ruiz, Santana_14_Queiroz}. To gain insight into the contribution of ectopic exchange or of any other transposition-independent mechanism to $\gamma_2$, let us consider the case of transposition-incompetent elements.

Though our genomes are saturated with transposons, the vast majority of them are no longer capable of transposition, having lost or damaged sequences critical to their replication \cite{Voliva_83_Edgell, Hardies_86_Edgell, Sassaman_97_Kazazian, Ostertag_01_Kazazian, Feschotte_02_Wessler, Brouha_03_Kazazian, Brookfield_05, Yang_09_Bennetzen_2}. If these ``dead'' transposons posed a meaningful threat to host fitness, then we would expect to see them quickly removed from the population. Their ``live'' cousins may also be deleterious, but at least they have the capacity to replicate to offset this shortcoming, and they should therefore be fitter and more abundant than their transposition-incompetent counterparts.

To make this argument more mathematically precise, let us decompose our simple model into ``live'' and ``dead'' populations:
\begin{align}
\frac{d \lambda_\text{live}}{dt} &= (\beta -\gamma_1) \lambda_\text{live} - \gamma_2 \lambda_\text{live} (\lambda_\text{live} + \lambda_\text{dead}) \\
\frac{d \lambda_\text{dead}}{dt} &= -\gamma_1 \lambda_\text{dead} - \gamma_2 \lambda_\text{dead} (\lambda_\text{live} + \lambda_\text{dead})
\end{align}

If $\lambda_\text{live} \ll \lambda_\text{dead} $, we may neglect its quadratic contribution to the $\gamma_2$ term, leaving us again with a linear differential equation with no equilibrium. Furthermore, since $\frac{d \lambda_\text{dead}}{dt} < \frac{d \lambda_\text{live}}{dt}$, the dead transposons will be whittled out of the population. Thus, the transposition-independent hypothesis is not consistent with the observed preponderance of transposition-incompetent elements.

Finally in all of these models, we would expect transposons to evolve to ever-increasing replication rates, which would appear to be possible by comparison with viral replication rates, but in nature we tend to see very low transposon replication rates \cite{Shen_87_Kleckner, Harada_90_Mukai, Nuzhdin_94_Mackay, Suh_95_Harada, Maside_00_Charlesworth, Sousa_13_Gordo}. On all of these grounds, we must reject the notion that nonlinear toxicities are relevant to stabilizing transposon populations.

\bibliography{references}

\begin{thebibliography}{60}%
\makeatletter
\providecommand \@ifxundefined [1]{%
 \@ifx{#1\undefined}
}%
\providecommand \@ifnum [1]{%
 \ifnum #1\expandafter \@firstoftwo
 \else \expandafter \@secondoftwo
 \fi
}%
\providecommand \@ifx [1]{%
 \ifx #1\expandafter \@firstoftwo
 \else \expandafter \@secondoftwo
 \fi
}%
\providecommand \natexlab [1]{#1}%
\providecommand \enquote  [1]{``#1''}%
\providecommand \bibnamefont  [1]{#1}%
\providecommand \bibfnamefont [1]{#1}%
\providecommand \citenamefont [1]{#1}%
\providecommand \href@noop [0]{\@secondoftwo}%
\providecommand \href [0]{\begingroup \@sanitize@url \@href}%
\providecommand \@href[1]{\@@startlink{#1}\@@href}%
\providecommand \@@href[1]{\endgroup#1\@@endlink}%
\providecommand \@sanitize@url [0]{\catcode `\\12\catcode `\$12\catcode `\&12\catcode `\#12\catcode `\^12\catcode `\_12\catcode `\%12\relax}%
\providecommand \@@startlink[1]{}%
\providecommand \@@endlink[0]{}%
\providecommand \url  [0]{\begingroup\@sanitize@url \@url }%
\providecommand \@url [1]{\endgroup\@href {#1}{\urlprefix }}%
\providecommand \urlprefix  [0]{URL }%
\providecommand \Eprint [0]{\href }%
\providecommand \doibase [0]{https://doi.org/}%
\providecommand \selectlanguage [0]{\@gobble}%
\providecommand \bibinfo  [0]{\@secondoftwo}%
\providecommand \bibfield  [0]{\@secondoftwo}%
\providecommand \translation [1]{[#1]}%
\providecommand \BibitemOpen [0]{}%
\providecommand \bibitemStop [0]{}%
\providecommand \bibitemNoStop [0]{.\EOS\space}%
\providecommand \EOS [0]{\spacefactor3000\relax}%
\providecommand \BibitemShut  [1]{\csname bibitem#1\endcsname}%
\let\auto@bib@innerbib\@empty
\bibitem [{\citenamefont {Siguier}\ \emph {et~al.}(2014)\citenamefont {Siguier}, \citenamefont {Gourbeyre},\ and\ \citenamefont {Chandler}}]{Siguier_14_Chandler}%
  \BibitemOpen
  \bibfield  {author} {\bibinfo {author} {\bibfnamefont {P.}~\bibnamefont {Siguier}}, \bibinfo {author} {\bibfnamefont {E.}~\bibnamefont {Gourbeyre}},\ and\ \bibinfo {author} {\bibfnamefont {M.}~\bibnamefont {Chandler}},\ }\bibfield  {title} {\bibinfo {title} {{Bacterial insertion sequences: their genomic impact and diversity}},\ }\href {https://doi.org/10.1111/1574-6976.12067} {\bibfield  {journal} {\bibinfo  {journal} {FEMS Microbiology Reviews}\ }\textbf {\bibinfo {volume} {38}},\ \bibinfo {pages} {865} (\bibinfo {year} {2014})},\ \Eprint {https://arxiv.org/abs/https://academic.oup.com/femsre/article-pdf/38/5/865/18142951/38-5-865.pdf} {https://academic.oup.com/femsre/article-pdf/38/5/865/18142951/38-5-865.pdf} \BibitemShut {NoStop}%
\bibitem [{\citenamefont {Thomas}\ and\ \citenamefont {Pritham}(2015)}]{Thomas_15_Pritham}%
  \BibitemOpen
  \bibfield  {author} {\bibinfo {author} {\bibfnamefont {J.}~\bibnamefont {Thomas}}\ and\ \bibinfo {author} {\bibfnamefont {E.~J.}\ \bibnamefont {Pritham}},\ }\bibfield  {title} {\bibinfo {title} {Helitrons, the eukaryotic rolling-circle transposable elements},\ }\href {https://doi.org/10.1128/microbiolspec.mdna3-0049-2014} {\bibfield  {journal} {\bibinfo  {journal} {Microbiology Spectrum}\ }\textbf {\bibinfo {volume} {3}},\ \bibinfo {pages} {10.1128/microbiolspec.mdna3} (\bibinfo {year} {2015})},\ \Eprint {https://arxiv.org/abs/https://journals.asm.org/doi/pdf/10.1128/microbiolspec.mdna3-0049-2014} {https://journals.asm.org/doi/pdf/10.1128/microbiolspec.mdna3-0049-2014} \BibitemShut {NoStop}%
\bibitem [{\citenamefont {Wells}\ and\ \citenamefont {Feschotte}(2020)}]{Wells_20_Feschotte}%
  \BibitemOpen
  \bibfield  {author} {\bibinfo {author} {\bibfnamefont {J.}~\bibnamefont {Wells}}\ and\ \bibinfo {author} {\bibfnamefont {C.}~\bibnamefont {Feschotte}},\ }\bibfield  {title} {\bibinfo {title} {A field guide to eukaryotic transposable elements},\ }\href {https://doi.org/10.1146/annurev-genet-040620-022145} {\bibfield  {journal} {\bibinfo  {journal} {Annu Rev Genet}\ }\textbf {\bibinfo {volume} {54}},\ \bibinfo {pages} {539} (\bibinfo {year} {2020})},\ \bibinfo {note} {pMID: 32955944}\BibitemShut {NoStop}%
\bibitem [{\citenamefont {Lander}\ \emph {et~al.}(2001)\citenamefont {Lander}, \citenamefont {Linton}, \citenamefont {Birren}, \citenamefont {Nusbaum}, \citenamefont {Zody}, \citenamefont {Baldwin}, \citenamefont {Devon}, \citenamefont {Dewar}, \citenamefont {Doyle}, \citenamefont {FitzHugh}, \citenamefont {Funke}, \citenamefont {Gage}, \citenamefont {Harris}, \citenamefont {Heaford}, \citenamefont {Howland}, \citenamefont {Kann}, \citenamefont {Lehoczky}, \citenamefont {LeVine}, \citenamefont {McEwan}, \citenamefont {McKernan}, \citenamefont {Meldrim}, \citenamefont {Mesirov}, \citenamefont {Miranda}, \citenamefont {Morris}, \citenamefont {Naylor}, \citenamefont {Raymond}, \citenamefont {Rosetti}, \citenamefont {Santos}, \citenamefont {Sheridan}, \citenamefont {Sougnez}, \citenamefont {Stange-Thomann}, \citenamefont {Stojanovic}, \citenamefont {Subramanian}, \citenamefont {Wyman}, \citenamefont {Rogers}, \citenamefont {Sulston}, \citenamefont {Ainscough}, \citenamefont {Beck}, \citenamefont {Bentley},
  \citenamefont {Burton}, \citenamefont {Clee}, \citenamefont {Carter}, \citenamefont {Coulson}, \citenamefont {Deadman}, \citenamefont {Deloukas}, \citenamefont {Dunham}, \citenamefont {Dunham}, \citenamefont {Durbin}, \citenamefont {French}, \citenamefont {Grafham}, \citenamefont {Gregory}, \citenamefont {Hubbard}, \citenamefont {Humphray}, \citenamefont {Hunt}, \citenamefont {Jones}, \citenamefont {Lloyd}, \citenamefont {McMurray}, \citenamefont {Matthews}, \citenamefont {Mercer}, \citenamefont {Milne}, \citenamefont {Mullikin}, \citenamefont {Mungall}, \citenamefont {Plumb}, \citenamefont {Ross}, \citenamefont {Shownkeen}, \citenamefont {Sims}, \citenamefont {Waterston}, \citenamefont {Wilson}, \citenamefont {Hillier}, \citenamefont {McPherson}, \citenamefont {Marra}, \citenamefont {Mardis}, \citenamefont {Fulton}, \citenamefont {Chinwalla}, \citenamefont {Pepin}, \citenamefont {Gish}, \citenamefont {Chissoe}, \citenamefont {Wendl}, \citenamefont {Delehaunty}, \citenamefont {Miner}, \citenamefont
  {Delehaunty}, \citenamefont {Kramer}, \citenamefont {Cook}, \citenamefont {Fulton}, \citenamefont {Johnson}, \citenamefont {Minx}, \citenamefont {Clifton}, \citenamefont {Hawkins}, \citenamefont {Branscomb}, \citenamefont {Predki}, \citenamefont {Richardson}, \citenamefont {Wenning}, \citenamefont {Slezak}, \citenamefont {Doggett}, \citenamefont {Cheng}, \citenamefont {Olsen}, \citenamefont {Lucas}, \citenamefont {Elkin}, \citenamefont {Uberbacher}, \citenamefont {Frazier}, \citenamefont {Gibbs}, \citenamefont {Muzny}, \citenamefont {Scherer}, \citenamefont {Bouck}, \citenamefont {Sodergren}, \citenamefont {Worley}, \citenamefont {Rives}, \citenamefont {Gorrell}, \citenamefont {Metzker}, \citenamefont {Naylor}, \citenamefont {Kucherlapati}, \citenamefont {Nelson}, \citenamefont {Weinstock}, \citenamefont {Sakaki}, \citenamefont {Fujiyama}, \citenamefont {Hattori}, \citenamefont {Yada}, \citenamefont {Toyoda}, \citenamefont {Itoh}, \citenamefont {Kawagoe}, \citenamefont {Watanabe}, \citenamefont {Totoki},
  \citenamefont {Taylor}, \citenamefont {Weissenbach}, \citenamefont {Heilig}, \citenamefont {Saurin}, \citenamefont {Artiguenave}, \citenamefont {Brottier}, \citenamefont {Bruls}, \citenamefont {Pelletier}, \citenamefont {Robert}, \citenamefont {Wincker}, \citenamefont {Smith}, \citenamefont {Doucette-Stamm}, \citenamefont {Rubenfield}, \citenamefont {Weinstock}, \citenamefont {Lee}, \citenamefont {Dubois}, \citenamefont {Rosenthal}, \citenamefont {Platzer}, \citenamefont {Nyakatura}, \citenamefont {Taudien}, \citenamefont {Rump}, \citenamefont {Yang}, \citenamefont {Yu}, \citenamefont {Wang}, \citenamefont {Huang}, \citenamefont {Gu}, \citenamefont {Hood}, \citenamefont {Rowen}, \citenamefont {Madan}, \citenamefont {Qin}, \citenamefont {Davis}, \citenamefont {Federspiel}, \citenamefont {Abola}, \citenamefont {Proctor}, \citenamefont {Myers}, \citenamefont {Schmutz}, \citenamefont {Dickson}, \citenamefont {Grimwood}, \citenamefont {Cox}, \citenamefont {Olson}, \citenamefont {Kaul}, \citenamefont {Raymond},
  \citenamefont {Shimizu}, \citenamefont {Kawasaki}, \citenamefont {Minoshima}, \citenamefont {Evans}, \citenamefont {Athanasiou}, \citenamefont {Schultz}, \citenamefont {Roe}, \citenamefont {Chen}, \citenamefont {Pan}, \citenamefont {Ramser}, \citenamefont {Lehrach}, \citenamefont {Reinhardt}, \citenamefont {McCombie}, \citenamefont {de~la Bastide}, \citenamefont {Dedhia}, \citenamefont {Blöcker}, \citenamefont {Hornischer}, \citenamefont {Nordsiek}, \citenamefont {Agarwala}, \citenamefont {Aravind}, \citenamefont {Bailey}, \citenamefont {Bateman}, \citenamefont {Batzoglou}, \citenamefont {Birney}, \citenamefont {Bork}, \citenamefont {Brown}, \citenamefont {Burge}, \citenamefont {Cerutti}, \citenamefont {Chen}, \citenamefont {Church}, \citenamefont {Clamp}, \citenamefont {Copley}, \citenamefont {Doerks}, \citenamefont {Eddy}, \citenamefont {Eichler}, \citenamefont {Furey}, \citenamefont {Galagan}, \citenamefont {Gilbert}, \citenamefont {Harmon}, \citenamefont {Hayashizaki}, \citenamefont {Haussler},
  \citenamefont {Hermjakob}, \citenamefont {Hokamp}, \citenamefont {Jang}, \citenamefont {Johnson}, \citenamefont {Jones}, \citenamefont {Kasif}, \citenamefont {Kaspryzk}, \citenamefont {Kennedy}, \citenamefont {Kent}, \citenamefont {Kitts}, \citenamefont {Koonin}, \citenamefont {Korf}, \citenamefont {Kulp}, \citenamefont {Lancet}, \citenamefont {Lowe}, \citenamefont {McLysaght}, \citenamefont {Mikkelsen}, \citenamefont {Moran}, \citenamefont {Mulder}, \citenamefont {Pollara}, \citenamefont {Ponting}, \citenamefont {Schuler}, \citenamefont {Schultz}, \citenamefont {Slater}, \citenamefont {Smit}, \citenamefont {Stupka}, \citenamefont {Szustakowki}, \citenamefont {Thierry-Mieg}, \citenamefont {Thierry-Mieg}, \citenamefont {Wagner}, \citenamefont {Wallis}, \citenamefont {Wheeler}, \citenamefont {Williams}, \citenamefont {Wolf}, \citenamefont {Wolfe}, \citenamefont {Yang}, \citenamefont {Yeh}, \citenamefont {Collins}, \citenamefont {Guyer}, \citenamefont {Peterson}, \citenamefont {Felsenfeld}, \citenamefont
  {Wetterstrand}, \citenamefont {Patrinos}, \citenamefont {Morgan}, \citenamefont {de}, \citenamefont {Catanese}, \citenamefont {Osoegawa}, \citenamefont {Shizuya}, \citenamefont {Choi}, \citenamefont {Chen},\ and\ \citenamefont {Szustakowki}}]{Lander_01}%
  \BibitemOpen
  \bibfield  {author} {\bibinfo {author} {\bibfnamefont {E.}~\bibnamefont {Lander}}, \bibinfo {author} {\bibfnamefont {L.}~\bibnamefont {Linton}}, \bibinfo {author} {\bibfnamefont {B.}~\bibnamefont {Birren}}, \bibinfo {author} {\bibfnamefont {C.}~\bibnamefont {Nusbaum}}, \bibinfo {author} {\bibfnamefont {M.}~\bibnamefont {Zody}}, \bibinfo {author} {\bibfnamefont {J.}~\bibnamefont {Baldwin}}, \bibinfo {author} {\bibfnamefont {K.}~\bibnamefont {Devon}}, \bibinfo {author} {\bibfnamefont {K.}~\bibnamefont {Dewar}}, \bibinfo {author} {\bibfnamefont {M.}~\bibnamefont {Doyle}}, \bibinfo {author} {\bibfnamefont {W.}~\bibnamefont {FitzHugh}}, \bibinfo {author} {\bibfnamefont {R.}~\bibnamefont {Funke}}, \bibinfo {author} {\bibfnamefont {D.}~\bibnamefont {Gage}}, \bibinfo {author} {\bibfnamefont {K.}~\bibnamefont {Harris}}, \bibinfo {author} {\bibfnamefont {A.}~\bibnamefont {Heaford}}, \bibinfo {author} {\bibfnamefont {J.}~\bibnamefont {Howland}}, \bibinfo {author} {\bibfnamefont {L.}~\bibnamefont {Kann}}, \bibinfo
  {author} {\bibfnamefont {J.}~\bibnamefont {Lehoczky}}, \bibinfo {author} {\bibfnamefont {R.}~\bibnamefont {LeVine}}, \bibinfo {author} {\bibfnamefont {P.}~\bibnamefont {McEwan}}, \bibinfo {author} {\bibfnamefont {K.}~\bibnamefont {McKernan}}, \bibinfo {author} {\bibfnamefont {J.}~\bibnamefont {Meldrim}}, \bibinfo {author} {\bibfnamefont {J.}~\bibnamefont {Mesirov}}, \bibinfo {author} {\bibfnamefont {C.}~\bibnamefont {Miranda}}, \bibinfo {author} {\bibfnamefont {W.}~\bibnamefont {Morris}}, \bibinfo {author} {\bibfnamefont {J.}~\bibnamefont {Naylor}}, \bibinfo {author} {\bibfnamefont {C.}~\bibnamefont {Raymond}}, \bibinfo {author} {\bibfnamefont {M.}~\bibnamefont {Rosetti}}, \bibinfo {author} {\bibfnamefont {R.}~\bibnamefont {Santos}}, \bibinfo {author} {\bibfnamefont {A.}~\bibnamefont {Sheridan}}, \bibinfo {author} {\bibfnamefont {C.}~\bibnamefont {Sougnez}}, \bibinfo {author} {\bibfnamefont {Y.}~\bibnamefont {Stange-Thomann}}, \bibinfo {author} {\bibfnamefont {N.}~\bibnamefont {Stojanovic}}, \bibinfo
  {author} {\bibfnamefont {A.}~\bibnamefont {Subramanian}}, \bibinfo {author} {\bibfnamefont {D.}~\bibnamefont {Wyman}}, \bibinfo {author} {\bibfnamefont {J.}~\bibnamefont {Rogers}}, \bibinfo {author} {\bibfnamefont {J.}~\bibnamefont {Sulston}}, \bibinfo {author} {\bibfnamefont {R.}~\bibnamefont {Ainscough}}, \bibinfo {author} {\bibfnamefont {S.}~\bibnamefont {Beck}}, \bibinfo {author} {\bibfnamefont {D.}~\bibnamefont {Bentley}}, \bibinfo {author} {\bibfnamefont {J.}~\bibnamefont {Burton}}, \bibinfo {author} {\bibfnamefont {C.}~\bibnamefont {Clee}}, \bibinfo {author} {\bibfnamefont {N.}~\bibnamefont {Carter}}, \bibinfo {author} {\bibfnamefont {A.}~\bibnamefont {Coulson}}, \bibinfo {author} {\bibfnamefont {R.}~\bibnamefont {Deadman}}, \bibinfo {author} {\bibfnamefont {P.}~\bibnamefont {Deloukas}}, \bibinfo {author} {\bibfnamefont {A.}~\bibnamefont {Dunham}}, \bibinfo {author} {\bibfnamefont {I.}~\bibnamefont {Dunham}}, \bibinfo {author} {\bibfnamefont {R.}~\bibnamefont {Durbin}}, \bibinfo {author}
  {\bibfnamefont {L.}~\bibnamefont {French}}, \bibinfo {author} {\bibfnamefont {D.}~\bibnamefont {Grafham}}, \bibinfo {author} {\bibfnamefont {S.}~\bibnamefont {Gregory}}, \bibinfo {author} {\bibfnamefont {T.}~\bibnamefont {Hubbard}}, \bibinfo {author} {\bibfnamefont {S.}~\bibnamefont {Humphray}}, \bibinfo {author} {\bibfnamefont {A.}~\bibnamefont {Hunt}}, \bibinfo {author} {\bibfnamefont {M.}~\bibnamefont {Jones}}, \bibinfo {author} {\bibfnamefont {C.}~\bibnamefont {Lloyd}}, \bibinfo {author} {\bibfnamefont {A.}~\bibnamefont {McMurray}}, \bibinfo {author} {\bibfnamefont {L.}~\bibnamefont {Matthews}}, \bibinfo {author} {\bibfnamefont {S.}~\bibnamefont {Mercer}}, \bibinfo {author} {\bibfnamefont {S.}~\bibnamefont {Milne}}, \bibinfo {author} {\bibfnamefont {J.}~\bibnamefont {Mullikin}}, \bibinfo {author} {\bibfnamefont {A.}~\bibnamefont {Mungall}}, \bibinfo {author} {\bibfnamefont {R.}~\bibnamefont {Plumb}}, \bibinfo {author} {\bibfnamefont {M.}~\bibnamefont {Ross}}, \bibinfo {author} {\bibfnamefont
  {R.}~\bibnamefont {Shownkeen}}, \bibinfo {author} {\bibfnamefont {S.}~\bibnamefont {Sims}}, \bibinfo {author} {\bibfnamefont {R.}~\bibnamefont {Waterston}}, \bibinfo {author} {\bibfnamefont {R.}~\bibnamefont {Wilson}}, \bibinfo {author} {\bibfnamefont {L.}~\bibnamefont {Hillier}}, \bibinfo {author} {\bibfnamefont {J.}~\bibnamefont {McPherson}}, \bibinfo {author} {\bibfnamefont {M.}~\bibnamefont {Marra}}, \bibinfo {author} {\bibfnamefont {E.}~\bibnamefont {Mardis}}, \bibinfo {author} {\bibfnamefont {L.}~\bibnamefont {Fulton}}, \bibinfo {author} {\bibfnamefont {A.}~\bibnamefont {Chinwalla}}, \bibinfo {author} {\bibfnamefont {K.}~\bibnamefont {Pepin}}, \bibinfo {author} {\bibfnamefont {W.}~\bibnamefont {Gish}}, \bibinfo {author} {\bibfnamefont {S.}~\bibnamefont {Chissoe}}, \bibinfo {author} {\bibfnamefont {M.}~\bibnamefont {Wendl}}, \bibinfo {author} {\bibfnamefont {K.}~\bibnamefont {Delehaunty}}, \bibinfo {author} {\bibfnamefont {T.}~\bibnamefont {Miner}}, \bibinfo {author} {\bibfnamefont {A.}~\bibnamefont
  {Delehaunty}}, \bibinfo {author} {\bibfnamefont {J.}~\bibnamefont {Kramer}}, \bibinfo {author} {\bibfnamefont {L.}~\bibnamefont {Cook}}, \bibinfo {author} {\bibfnamefont {R.}~\bibnamefont {Fulton}}, \bibinfo {author} {\bibfnamefont {D.}~\bibnamefont {Johnson}}, \bibinfo {author} {\bibfnamefont {P.}~\bibnamefont {Minx}}, \bibinfo {author} {\bibfnamefont {S.}~\bibnamefont {Clifton}}, \bibinfo {author} {\bibfnamefont {T.}~\bibnamefont {Hawkins}}, \bibinfo {author} {\bibfnamefont {E.}~\bibnamefont {Branscomb}}, \bibinfo {author} {\bibfnamefont {P.}~\bibnamefont {Predki}}, \bibinfo {author} {\bibfnamefont {P.}~\bibnamefont {Richardson}}, \bibinfo {author} {\bibfnamefont {S.}~\bibnamefont {Wenning}}, \bibinfo {author} {\bibfnamefont {T.}~\bibnamefont {Slezak}}, \bibinfo {author} {\bibfnamefont {N.}~\bibnamefont {Doggett}}, \bibinfo {author} {\bibfnamefont {J.}~\bibnamefont {Cheng}}, \bibinfo {author} {\bibfnamefont {A.}~\bibnamefont {Olsen}}, \bibinfo {author} {\bibfnamefont {S.}~\bibnamefont {Lucas}}, \bibinfo
  {author} {\bibfnamefont {C.}~\bibnamefont {Elkin}}, \bibinfo {author} {\bibfnamefont {E.}~\bibnamefont {Uberbacher}}, \bibinfo {author} {\bibfnamefont {M.}~\bibnamefont {Frazier}}, \bibinfo {author} {\bibfnamefont {R.}~\bibnamefont {Gibbs}}, \bibinfo {author} {\bibfnamefont {D.}~\bibnamefont {Muzny}}, \bibinfo {author} {\bibfnamefont {S.}~\bibnamefont {Scherer}}, \bibinfo {author} {\bibfnamefont {J.}~\bibnamefont {Bouck}}, \bibinfo {author} {\bibfnamefont {E.}~\bibnamefont {Sodergren}}, \bibinfo {author} {\bibfnamefont {K.}~\bibnamefont {Worley}}, \bibinfo {author} {\bibfnamefont {C.}~\bibnamefont {Rives}}, \bibinfo {author} {\bibfnamefont {J.}~\bibnamefont {Gorrell}}, \bibinfo {author} {\bibfnamefont {M.}~\bibnamefont {Metzker}}, \bibinfo {author} {\bibfnamefont {S.}~\bibnamefont {Naylor}}, \bibinfo {author} {\bibfnamefont {R.}~\bibnamefont {Kucherlapati}}, \bibinfo {author} {\bibfnamefont {D.}~\bibnamefont {Nelson}}, \bibinfo {author} {\bibfnamefont {G.}~\bibnamefont {Weinstock}}, \bibinfo {author}
  {\bibfnamefont {Y.}~\bibnamefont {Sakaki}}, \bibinfo {author} {\bibfnamefont {A.}~\bibnamefont {Fujiyama}}, \bibinfo {author} {\bibfnamefont {M.}~\bibnamefont {Hattori}}, \bibinfo {author} {\bibfnamefont {T.}~\bibnamefont {Yada}}, \bibinfo {author} {\bibfnamefont {A.}~\bibnamefont {Toyoda}}, \bibinfo {author} {\bibfnamefont {T.}~\bibnamefont {Itoh}}, \bibinfo {author} {\bibfnamefont {C.}~\bibnamefont {Kawagoe}}, \bibinfo {author} {\bibfnamefont {H.}~\bibnamefont {Watanabe}}, \bibinfo {author} {\bibfnamefont {Y.}~\bibnamefont {Totoki}}, \bibinfo {author} {\bibfnamefont {T.}~\bibnamefont {Taylor}}, \bibinfo {author} {\bibfnamefont {J.}~\bibnamefont {Weissenbach}}, \bibinfo {author} {\bibfnamefont {R.}~\bibnamefont {Heilig}}, \bibinfo {author} {\bibfnamefont {W.}~\bibnamefont {Saurin}}, \bibinfo {author} {\bibfnamefont {F.}~\bibnamefont {Artiguenave}}, \bibinfo {author} {\bibfnamefont {P.}~\bibnamefont {Brottier}}, \bibinfo {author} {\bibfnamefont {T.}~\bibnamefont {Bruls}}, \bibinfo {author} {\bibfnamefont
  {E.}~\bibnamefont {Pelletier}}, \bibinfo {author} {\bibfnamefont {C.}~\bibnamefont {Robert}}, \bibinfo {author} {\bibfnamefont {P.}~\bibnamefont {Wincker}}, \bibinfo {author} {\bibfnamefont {D.}~\bibnamefont {Smith}}, \bibinfo {author} {\bibfnamefont {L.}~\bibnamefont {Doucette-Stamm}}, \bibinfo {author} {\bibfnamefont {M.}~\bibnamefont {Rubenfield}}, \bibinfo {author} {\bibfnamefont {K.}~\bibnamefont {Weinstock}}, \bibinfo {author} {\bibfnamefont {H.}~\bibnamefont {Lee}}, \bibinfo {author} {\bibfnamefont {J.}~\bibnamefont {Dubois}}, \bibinfo {author} {\bibfnamefont {A.}~\bibnamefont {Rosenthal}}, \bibinfo {author} {\bibfnamefont {M.}~\bibnamefont {Platzer}}, \bibinfo {author} {\bibfnamefont {G.}~\bibnamefont {Nyakatura}}, \bibinfo {author} {\bibfnamefont {S.}~\bibnamefont {Taudien}}, \bibinfo {author} {\bibfnamefont {A.}~\bibnamefont {Rump}}, \bibinfo {author} {\bibfnamefont {H.}~\bibnamefont {Yang}}, \bibinfo {author} {\bibfnamefont {J.}~\bibnamefont {Yu}}, \bibinfo {author} {\bibfnamefont
  {J.}~\bibnamefont {Wang}}, \bibinfo {author} {\bibfnamefont {G.}~\bibnamefont {Huang}}, \bibinfo {author} {\bibfnamefont {J.}~\bibnamefont {Gu}}, \bibinfo {author} {\bibfnamefont {L.}~\bibnamefont {Hood}}, \bibinfo {author} {\bibfnamefont {L.}~\bibnamefont {Rowen}}, \bibinfo {author} {\bibfnamefont {A.}~\bibnamefont {Madan}}, \bibinfo {author} {\bibfnamefont {S.}~\bibnamefont {Qin}}, \bibinfo {author} {\bibfnamefont {R.}~\bibnamefont {Davis}}, \bibinfo {author} {\bibfnamefont {N.}~\bibnamefont {Federspiel}}, \bibinfo {author} {\bibfnamefont {A.}~\bibnamefont {Abola}}, \bibinfo {author} {\bibfnamefont {M.}~\bibnamefont {Proctor}}, \bibinfo {author} {\bibfnamefont {R.}~\bibnamefont {Myers}}, \bibinfo {author} {\bibfnamefont {J.}~\bibnamefont {Schmutz}}, \bibinfo {author} {\bibfnamefont {M.}~\bibnamefont {Dickson}}, \bibinfo {author} {\bibfnamefont {J.}~\bibnamefont {Grimwood}}, \bibinfo {author} {\bibfnamefont {D.}~\bibnamefont {Cox}}, \bibinfo {author} {\bibfnamefont {M.}~\bibnamefont {Olson}}, \bibinfo
  {author} {\bibfnamefont {R.}~\bibnamefont {Kaul}}, \bibinfo {author} {\bibfnamefont {C.}~\bibnamefont {Raymond}}, \bibinfo {author} {\bibfnamefont {N.}~\bibnamefont {Shimizu}}, \bibinfo {author} {\bibfnamefont {K.}~\bibnamefont {Kawasaki}}, \bibinfo {author} {\bibfnamefont {S.}~\bibnamefont {Minoshima}}, \bibinfo {author} {\bibfnamefont {G.}~\bibnamefont {Evans}}, \bibinfo {author} {\bibfnamefont {M.}~\bibnamefont {Athanasiou}}, \bibinfo {author} {\bibfnamefont {R.}~\bibnamefont {Schultz}}, \bibinfo {author} {\bibfnamefont {B.}~\bibnamefont {Roe}}, \bibinfo {author} {\bibfnamefont {F.}~\bibnamefont {Chen}}, \bibinfo {author} {\bibfnamefont {H.}~\bibnamefont {Pan}}, \bibinfo {author} {\bibfnamefont {J.}~\bibnamefont {Ramser}}, \bibinfo {author} {\bibfnamefont {H.}~\bibnamefont {Lehrach}}, \bibinfo {author} {\bibfnamefont {R.}~\bibnamefont {Reinhardt}}, \bibinfo {author} {\bibfnamefont {W.}~\bibnamefont {McCombie}}, \bibinfo {author} {\bibfnamefont {M.}~\bibnamefont {de~la Bastide}}, \bibinfo {author}
  {\bibfnamefont {N.}~\bibnamefont {Dedhia}}, \bibinfo {author} {\bibfnamefont {H.}~\bibnamefont {Blöcker}}, \bibinfo {author} {\bibfnamefont {K.}~\bibnamefont {Hornischer}}, \bibinfo {author} {\bibfnamefont {G.}~\bibnamefont {Nordsiek}}, \bibinfo {author} {\bibfnamefont {R.}~\bibnamefont {Agarwala}}, \bibinfo {author} {\bibfnamefont {L.}~\bibnamefont {Aravind}}, \bibinfo {author} {\bibfnamefont {J.}~\bibnamefont {Bailey}}, \bibinfo {author} {\bibfnamefont {A.}~\bibnamefont {Bateman}}, \bibinfo {author} {\bibfnamefont {S.}~\bibnamefont {Batzoglou}}, \bibinfo {author} {\bibfnamefont {E.}~\bibnamefont {Birney}}, \bibinfo {author} {\bibfnamefont {P.}~\bibnamefont {Bork}}, \bibinfo {author} {\bibfnamefont {D.}~\bibnamefont {Brown}}, \bibinfo {author} {\bibfnamefont {C.}~\bibnamefont {Burge}}, \bibinfo {author} {\bibfnamefont {L.}~\bibnamefont {Cerutti}}, \bibinfo {author} {\bibfnamefont {H.}~\bibnamefont {Chen}}, \bibinfo {author} {\bibfnamefont {D.}~\bibnamefont {Church}}, \bibinfo {author} {\bibfnamefont
  {M.}~\bibnamefont {Clamp}}, \bibinfo {author} {\bibfnamefont {R.}~\bibnamefont {Copley}}, \bibinfo {author} {\bibfnamefont {T.}~\bibnamefont {Doerks}}, \bibinfo {author} {\bibfnamefont {S.}~\bibnamefont {Eddy}}, \bibinfo {author} {\bibfnamefont {E.}~\bibnamefont {Eichler}}, \bibinfo {author} {\bibfnamefont {T.}~\bibnamefont {Furey}}, \bibinfo {author} {\bibfnamefont {J.}~\bibnamefont {Galagan}}, \bibinfo {author} {\bibfnamefont {J.}~\bibnamefont {Gilbert}}, \bibinfo {author} {\bibfnamefont {C.}~\bibnamefont {Harmon}}, \bibinfo {author} {\bibfnamefont {Y.}~\bibnamefont {Hayashizaki}}, \bibinfo {author} {\bibfnamefont {D.}~\bibnamefont {Haussler}}, \bibinfo {author} {\bibfnamefont {H.}~\bibnamefont {Hermjakob}}, \bibinfo {author} {\bibfnamefont {K.}~\bibnamefont {Hokamp}}, \bibinfo {author} {\bibfnamefont {W.}~\bibnamefont {Jang}}, \bibinfo {author} {\bibfnamefont {L.}~\bibnamefont {Johnson}}, \bibinfo {author} {\bibfnamefont {T.}~\bibnamefont {Jones}}, \bibinfo {author} {\bibfnamefont {S.}~\bibnamefont
  {Kasif}}, \bibinfo {author} {\bibfnamefont {A.}~\bibnamefont {Kaspryzk}}, \bibinfo {author} {\bibfnamefont {S.}~\bibnamefont {Kennedy}}, \bibinfo {author} {\bibfnamefont {W.}~\bibnamefont {Kent}}, \bibinfo {author} {\bibfnamefont {P.}~\bibnamefont {Kitts}}, \bibinfo {author} {\bibfnamefont {E.}~\bibnamefont {Koonin}}, \bibinfo {author} {\bibfnamefont {I.}~\bibnamefont {Korf}}, \bibinfo {author} {\bibfnamefont {D.}~\bibnamefont {Kulp}}, \bibinfo {author} {\bibfnamefont {D.}~\bibnamefont {Lancet}}, \bibinfo {author} {\bibfnamefont {T.}~\bibnamefont {Lowe}}, \bibinfo {author} {\bibfnamefont {A.}~\bibnamefont {McLysaght}}, \bibinfo {author} {\bibfnamefont {T.}~\bibnamefont {Mikkelsen}}, \bibinfo {author} {\bibfnamefont {J.}~\bibnamefont {Moran}}, \bibinfo {author} {\bibfnamefont {N.}~\bibnamefont {Mulder}}, \bibinfo {author} {\bibfnamefont {V.}~\bibnamefont {Pollara}}, \bibinfo {author} {\bibfnamefont {C.}~\bibnamefont {Ponting}}, \bibinfo {author} {\bibfnamefont {G.}~\bibnamefont {Schuler}}, \bibinfo {author}
  {\bibfnamefont {J.}~\bibnamefont {Schultz}}, \bibinfo {author} {\bibfnamefont {G.}~\bibnamefont {Slater}}, \bibinfo {author} {\bibfnamefont {A.}~\bibnamefont {Smit}}, \bibinfo {author} {\bibfnamefont {E.}~\bibnamefont {Stupka}}, \bibinfo {author} {\bibfnamefont {J.}~\bibnamefont {Szustakowki}}, \bibinfo {author} {\bibfnamefont {D.}~\bibnamefont {Thierry-Mieg}}, \bibinfo {author} {\bibfnamefont {J.}~\bibnamefont {Thierry-Mieg}}, \bibinfo {author} {\bibfnamefont {L.}~\bibnamefont {Wagner}}, \bibinfo {author} {\bibfnamefont {J.}~\bibnamefont {Wallis}}, \bibinfo {author} {\bibfnamefont {R.}~\bibnamefont {Wheeler}}, \bibinfo {author} {\bibfnamefont {A.}~\bibnamefont {Williams}}, \bibinfo {author} {\bibfnamefont {Y.}~\bibnamefont {Wolf}}, \bibinfo {author} {\bibfnamefont {K.}~\bibnamefont {Wolfe}}, \bibinfo {author} {\bibfnamefont {S.}~\bibnamefont {Yang}}, \bibinfo {author} {\bibfnamefont {R.}~\bibnamefont {Yeh}}, \bibinfo {author} {\bibfnamefont {F.}~\bibnamefont {Collins}}, \bibinfo {author} {\bibfnamefont
  {M.}~\bibnamefont {Guyer}}, \bibinfo {author} {\bibfnamefont {J.}~\bibnamefont {Peterson}}, \bibinfo {author} {\bibfnamefont {A.}~\bibnamefont {Felsenfeld}}, \bibinfo {author} {\bibfnamefont {K.}~\bibnamefont {Wetterstrand}}, \bibinfo {author} {\bibfnamefont {A.}~\bibnamefont {Patrinos}}, \bibinfo {author} {\bibfnamefont {M.}~\bibnamefont {Morgan}}, \bibinfo {author} {\bibfnamefont {P.}~\bibnamefont {de}, \bibfnamefont {Jong}}, \bibinfo {author} {\bibfnamefont {J.}~\bibnamefont {Catanese}}, \bibinfo {author} {\bibfnamefont {K.}~\bibnamefont {Osoegawa}}, \bibinfo {author} {\bibfnamefont {H.}~\bibnamefont {Shizuya}}, \bibinfo {author} {\bibfnamefont {S.}~\bibnamefont {Choi}}, \bibinfo {author} {\bibfnamefont {Y.}~\bibnamefont {Chen}},\ and\ \bibinfo {author} {\bibfnamefont {J.}~\bibnamefont {Szustakowki}},\ }\bibfield  {title} {\bibinfo {title} {Initial sequencing and analysis of the human genome},\ }\href {https://doi.org/10.1038/35057062} {\bibfield  {journal} {\bibinfo  {journal} {Nature}\ }\textbf
  {\bibinfo {volume} {409(6822)}},\ \bibinfo {pages} {860} (\bibinfo {year} {2001})},\ \bibinfo {note} {pMID: 11237011}\BibitemShut {NoStop}%
\bibitem [{\citenamefont {Kazazian}\ \emph {et~al.}(1988)\citenamefont {Kazazian}, \citenamefont {Wong}, \citenamefont {Youssoufian}, \citenamefont {Scott}, \citenamefont {Phillips},\ and\ \citenamefont {Antonarakis}}]{Kazazian_88_Antonarakis}%
  \BibitemOpen
  \bibfield  {author} {\bibinfo {author} {\bibfnamefont {J.}~\bibnamefont {Kazazian}, \bibfnamefont {HH}}, \bibinfo {author} {\bibfnamefont {C.}~\bibnamefont {Wong}}, \bibinfo {author} {\bibfnamefont {H.}~\bibnamefont {Youssoufian}}, \bibinfo {author} {\bibfnamefont {A.}~\bibnamefont {Scott}}, \bibinfo {author} {\bibfnamefont {D.}~\bibnamefont {Phillips}},\ and\ \bibinfo {author} {\bibfnamefont {S.}~\bibnamefont {Antonarakis}},\ }\bibfield  {title} {\bibinfo {title} {Haemophilia a resulting from de novo insertion of l1 sequences represents a novel mechanism for mutation in man},\ }\href {https://doi.org/10.1038/332164a0} {\bibfield  {journal} {\bibinfo  {journal} {Nature}\ }\textbf {\bibinfo {volume} {332(6160)}},\ \bibinfo {pages} {164} (\bibinfo {year} {1988})},\ \bibinfo {note} {pMID: 2831458}\BibitemShut {NoStop}%
\bibitem [{\citenamefont {Pradhan}\ and\ \citenamefont {Ramakrishna}(2022)}]{Pradhan_22_Ramakrishna}%
  \BibitemOpen
  \bibfield  {author} {\bibinfo {author} {\bibfnamefont {R.~K.}\ \bibnamefont {Pradhan}}\ and\ \bibinfo {author} {\bibfnamefont {W.}~\bibnamefont {Ramakrishna}},\ }\bibfield  {title} {\bibinfo {title} {Transposons: Unexpected players in cancer},\ }\href {https://doi.org/https://doi.org/10.1016/j.gene.2021.145975} {\bibfield  {journal} {\bibinfo  {journal} {Gene}\ }\textbf {\bibinfo {volume} {808}},\ \bibinfo {pages} {145975} (\bibinfo {year} {2022})}\BibitemShut {NoStop}%
\bibitem [{\citenamefont {Prak}\ and\ \citenamefont {Kazazian}(2000)}]{Prak_00_Kazazian}%
  \BibitemOpen
  \bibfield  {author} {\bibinfo {author} {\bibfnamefont {E.}~\bibnamefont {Prak}}\ and\ \bibinfo {author} {\bibfnamefont {H.~J.}\ \bibnamefont {Kazazian}},\ }\bibfield  {title} {\bibinfo {title} {Mobile elements and the human genome},\ }\href {https://doi.org/10.1038/35038572} {\bibfield  {journal} {\bibinfo  {journal} {Nat Rev Genet}\ }\textbf {\bibinfo {volume} {1(2)}},\ \bibinfo {pages} {134} (\bibinfo {year} {2000})},\ \bibinfo {note} {pMID: 11253653}\BibitemShut {NoStop}%
\bibitem [{\citenamefont {Havecker}\ \emph {et~al.}(2004)\citenamefont {Havecker}, \citenamefont {Gao},\ and\ \citenamefont {Voytas}}]{Havecker_04_Voytas}%
  \BibitemOpen
  \bibfield  {author} {\bibinfo {author} {\bibfnamefont {E.}~\bibnamefont {Havecker}}, \bibinfo {author} {\bibfnamefont {X.}~\bibnamefont {Gao}},\ and\ \bibinfo {author} {\bibfnamefont {D.}~\bibnamefont {Voytas}},\ }\bibfield  {title} {\bibinfo {title} {The diversity of ltr retrotransposons},\ }\href {https://doi.org/10.1186/gb-2004-5-6-225} {\bibfield  {journal} {\bibinfo  {journal} {Genome Biol}\ }\textbf {\bibinfo {volume} {5(6)}},\ \bibinfo {pages} {225} (\bibinfo {year} {2004})},\ \bibinfo {note} {pMID: 15186483}\BibitemShut {NoStop}%
\bibitem [{\citenamefont {Venkatesh}\ and\ \citenamefont {Nandini}(2020)}]{Venkatesh_20_Nandini}%
  \BibitemOpen
  \bibfield  {author} {\bibinfo {author} {\bibnamefont {Venkatesh}}\ and\ \bibinfo {author} {\bibfnamefont {B.}~\bibnamefont {Nandini}},\ }\bibfield  {title} {\bibinfo {title} {Miniature inverted-repeat transposable elements (mites), derived insertional polymorphism as a tool of marker systems for molecular plant breeding},\ }\href {https://doi.org/10.1007/s11033-020-05365-y} {\bibfield  {journal} {\bibinfo  {journal} {Mol Biol Rep}\ }\textbf {\bibinfo {volume} {47(4)}},\ \bibinfo {pages} {3155} (\bibinfo {year} {2020})},\ \bibinfo {note} {pMID: 32162128}\BibitemShut {NoStop}%
\bibitem [{\citenamefont {Leonardo}\ and\ \citenamefont {Nuzhdin}(2002)}]{Leonardo_02_Nuzhdin}%
  \BibitemOpen
  \bibfield  {author} {\bibinfo {author} {\bibfnamefont {T.}~\bibnamefont {Leonardo}}\ and\ \bibinfo {author} {\bibfnamefont {S.}~\bibnamefont {Nuzhdin}},\ }\bibfield  {title} {\bibinfo {title} {Intracellular battlegrounds: conflict and cooperation between transposable elements},\ }\href {https://doi.org/10.1017/s0016672302009710} {\bibfield  {journal} {\bibinfo  {journal} {Genet Res}\ }\textbf {\bibinfo {volume} {80(3)}},\ \bibinfo {pages} {155} (\bibinfo {year} {2002})},\ \bibinfo {note} {pMID: 12688654}\BibitemShut {NoStop}%
\bibitem [{\citenamefont {Brookfield}(2005)}]{Brookfield_05}%
  \BibitemOpen
  \bibfield  {author} {\bibinfo {author} {\bibfnamefont {J.}~\bibnamefont {Brookfield}},\ }\bibfield  {title} {\bibinfo {title} {The ecology of the genome - mobile dna elements and their hosts},\ }\href {https://doi.org/10.1038/nrg1524} {\bibfield  {journal} {\bibinfo  {journal} {Nat Rev Genet}\ }\textbf {\bibinfo {volume} {6(2)}},\ \bibinfo {pages} {128} (\bibinfo {year} {2005})},\ \bibinfo {note} {pMID: 15640810}\BibitemShut {NoStop}%
\bibitem [{\citenamefont {Kaplan}\ \emph {et~al.}(1985)\citenamefont {Kaplan}, \citenamefont {Darden},\ and\ \citenamefont {Langley}}]{Kaplan_85_Langley}%
  \BibitemOpen
  \bibfield  {author} {\bibinfo {author} {\bibfnamefont {N.}~\bibnamefont {Kaplan}}, \bibinfo {author} {\bibfnamefont {T.}~\bibnamefont {Darden}},\ and\ \bibinfo {author} {\bibfnamefont {C.}~\bibnamefont {Langley}},\ }\bibfield  {title} {\bibinfo {title} {Evolution and extinction of transposable elements in mendelian populations},\ }\href {https://doi.org/10.1093/genetics/109.2.459} {\bibfield  {journal} {\bibinfo  {journal} {Genetics}\ }\textbf {\bibinfo {volume} {109(2)}},\ \bibinfo {pages} {459} (\bibinfo {year} {1985})},\ \bibinfo {note} {pMID: 2982700}\BibitemShut {NoStop}%
\bibitem [{\citenamefont {Brookfield}(1991)}]{Brookfield_91}%
  \BibitemOpen
  \bibfield  {author} {\bibinfo {author} {\bibfnamefont {J.}~\bibnamefont {Brookfield}},\ }\bibfield  {title} {\bibinfo {title} {Models of repression of transposition in p-m hybrid dysgenesis by p cytotype and by zygotically encoded repressor proteins},\ }\href {https://doi.org/10.1093/genetics/128.2.471} {\bibfield  {journal} {\bibinfo  {journal} {Genetics}\ }\textbf {\bibinfo {volume} {128(2)}},\ \bibinfo {pages} {471} (\bibinfo {year} {1991})},\ \bibinfo {note} {pMID: 1649073}\BibitemShut {NoStop}%
\bibitem [{\citenamefont {Brookfield}(1996)}]{Brookfield_96}%
  \BibitemOpen
  \bibfield  {author} {\bibinfo {author} {\bibfnamefont {J.}~\bibnamefont {Brookfield}},\ }\bibfield  {title} {\bibinfo {title} {Models of the spread of non-autonomous selfish transposable elements when transposition and fitness are coupled},\ }\href {https://doi.org/10.1017/S0016672300033681} {\bibfield  {journal} {\bibinfo  {journal} {Genetics Research}\ }\textbf {\bibinfo {volume} {67(3)}},\ \bibinfo {pages} {199} (\bibinfo {year} {1996})}\BibitemShut {NoStop}%
\bibitem [{\citenamefont {Le}\ and\ \citenamefont {Capy}(2006)}]{LeRouzic_06_Capy}%
  \BibitemOpen
  \bibfield  {author} {\bibinfo {author} {\bibfnamefont {A.}~\bibnamefont {Le}, \bibfnamefont {Rouzic}}\ and\ \bibinfo {author} {\bibfnamefont {P.}~\bibnamefont {Capy}},\ }\bibfield  {title} {\bibinfo {title} {Population genetics models of competition between transposable element subfamilies},\ }\href {https://doi.org/10.1534/genetics.105.052241} {\bibfield  {journal} {\bibinfo  {journal} {Genetics}\ }\textbf {\bibinfo {volume} {174(2)}},\ \bibinfo {pages} {785} (\bibinfo {year} {2006})},\ \bibinfo {note} {pMID: 16888345}\BibitemShut {NoStop}%
\bibitem [{\citenamefont {Le}\ \emph {et~al.}(2007)\citenamefont {Le}, \citenamefont {Boutin},\ and\ \citenamefont {Capy}}]{LeRouzic_07_Capy}%
  \BibitemOpen
  \bibfield  {author} {\bibinfo {author} {\bibfnamefont {A.}~\bibnamefont {Le}, \bibfnamefont {Rouzic}}, \bibinfo {author} {\bibfnamefont {T.}~\bibnamefont {Boutin}},\ and\ \bibinfo {author} {\bibfnamefont {P.}~\bibnamefont {Capy}},\ }\bibfield  {title} {\bibinfo {title} {Long-term evolution of transposable elements},\ }\href {https://doi.org/10.1073/pnas.0705238104} {\bibfield  {journal} {\bibinfo  {journal} {Proc Natl Acad Sci U S A}\ }\textbf {\bibinfo {volume} {104(49)}},\ \bibinfo {pages} {19375} (\bibinfo {year} {2007})},\ \bibinfo {note} {pMID: 200}\BibitemShut {NoStop}%
\bibitem [{\citenamefont {Xue}\ and\ \citenamefont {Goldenfeld}(2016)}]{Xue_16_Goldenfeld}%
  \BibitemOpen
  \bibfield  {author} {\bibinfo {author} {\bibfnamefont {C.}~\bibnamefont {Xue}}\ and\ \bibinfo {author} {\bibfnamefont {N.}~\bibnamefont {Goldenfeld}},\ }\bibfield  {title} {\bibinfo {title} {Stochastic predator-prey dynamics of transposons in the human genome},\ }\href {https://doi.org/10.1103/PhysRevLett.117.208101} {\bibfield  {journal} {\bibinfo  {journal} {Phys Rev Lett}\ }\textbf {\bibinfo {volume} {117(20)}},\ \bibinfo {pages} {208101} (\bibinfo {year} {2016})},\ \bibinfo {note} {pMID: 27886494}\BibitemShut {NoStop}%
\bibitem [{\citenamefont {Feschotte}\ \emph {et~al.}(2002)\citenamefont {Feschotte}, \citenamefont {Jiang},\ and\ \citenamefont {Wessler}}]{Feschotte_02_Wessler}%
  \BibitemOpen
  \bibfield  {author} {\bibinfo {author} {\bibfnamefont {C.}~\bibnamefont {Feschotte}}, \bibinfo {author} {\bibfnamefont {N.}~\bibnamefont {Jiang}},\ and\ \bibinfo {author} {\bibfnamefont {S.}~\bibnamefont {Wessler}},\ }\bibfield  {title} {\bibinfo {title} {Plant transposable elements: where genetics meets genomics},\ }\href {https://doi.org/10.1038/nrg793} {\bibfield  {journal} {\bibinfo  {journal} {Nat Rev Genet}\ }\textbf {\bibinfo {volume} {3(5)}},\ \bibinfo {pages} {329} (\bibinfo {year} {2002})},\ \bibinfo {note} {pMID: 11988759}\BibitemShut {NoStop}%
\bibitem [{\citenamefont {Oggenfuss}\ \emph {et~al.}(2021)\citenamefont {Oggenfuss}, \citenamefont {Badet}, \citenamefont {Wicker}, \citenamefont {Hartmann}, \citenamefont {Singh}, \citenamefont {Abraham}, \citenamefont {Karisto}, \citenamefont {Vonlanthen}, \citenamefont {Mundt}, \citenamefont {McDonald},\ and\ \citenamefont {Croll}}]{Oggenfuss_21_Croll}%
  \BibitemOpen
  \bibfield  {author} {\bibinfo {author} {\bibfnamefont {U.}~\bibnamefont {Oggenfuss}}, \bibinfo {author} {\bibfnamefont {T.}~\bibnamefont {Badet}}, \bibinfo {author} {\bibfnamefont {T.}~\bibnamefont {Wicker}}, \bibinfo {author} {\bibfnamefont {F.~E.}\ \bibnamefont {Hartmann}}, \bibinfo {author} {\bibfnamefont {N.~K.}\ \bibnamefont {Singh}}, \bibinfo {author} {\bibfnamefont {L.}~\bibnamefont {Abraham}}, \bibinfo {author} {\bibfnamefont {P.}~\bibnamefont {Karisto}}, \bibinfo {author} {\bibfnamefont {T.}~\bibnamefont {Vonlanthen}}, \bibinfo {author} {\bibfnamefont {C.}~\bibnamefont {Mundt}}, \bibinfo {author} {\bibfnamefont {B.~A.}\ \bibnamefont {McDonald}},\ and\ \bibinfo {author} {\bibfnamefont {D.}~\bibnamefont {Croll}},\ }\bibfield  {title} {\bibinfo {title} {A population-level invasion by transposable elements triggers genome expansion in a fungal pathogen},\ }\href {https://doi.org/10.7554/eLife.69249} {\bibfield  {journal} {\bibinfo  {journal} {eLife}\ }\textbf {\bibinfo {volume} {10}},\ \bibinfo {pages}
  {e69249} (\bibinfo {year} {2021})}\BibitemShut {NoStop}%
\bibitem [{\citenamefont {Barro-Trastoy}\ and\ \citenamefont {Köhler}(2024)}]{BarroTrastoy_24_Kohler}%
  \BibitemOpen
  \bibfield  {author} {\bibinfo {author} {\bibfnamefont {D.}~\bibnamefont {Barro-Trastoy}}\ and\ \bibinfo {author} {\bibfnamefont {C.}~\bibnamefont {Köhler}},\ }\bibfield  {title} {\bibinfo {title} {Helitrons: genomic parasites that generate developmental novelties},\ }\href {https://doi.org/10.1016/j.tig.2024.02.002} {\bibfield  {journal} {\bibinfo  {journal} {Trends Genet}\ }\textbf {\bibinfo {volume} {40(5)}},\ \bibinfo {pages} {437} (\bibinfo {year} {2024})},\ \bibinfo {note} {pMID: 38429198}\BibitemShut {NoStop}%
\bibitem [{\citenamefont {Grabundzija}\ \emph {et~al.}(2016)\citenamefont {Grabundzija}, \citenamefont {Messing}, \citenamefont {Thomas}, \citenamefont {Cosby}, \citenamefont {Bilic}, \citenamefont {Miskey}, \citenamefont {Gogol-Döring}, \citenamefont {Kapitonov}, \citenamefont {Diem}, \citenamefont {Dalda}, \citenamefont {Jurka}, \citenamefont {Pritham}, \citenamefont {Dyda}, \citenamefont {Izsvák},\ and\ \citenamefont {Ivics}}]{Grabundzija_16_Ivics}%
  \BibitemOpen
  \bibfield  {author} {\bibinfo {author} {\bibfnamefont {I.}~\bibnamefont {Grabundzija}}, \bibinfo {author} {\bibfnamefont {S.}~\bibnamefont {Messing}}, \bibinfo {author} {\bibfnamefont {J.}~\bibnamefont {Thomas}}, \bibinfo {author} {\bibfnamefont {R.}~\bibnamefont {Cosby}}, \bibinfo {author} {\bibfnamefont {I.}~\bibnamefont {Bilic}}, \bibinfo {author} {\bibfnamefont {C.}~\bibnamefont {Miskey}}, \bibinfo {author} {\bibfnamefont {A.}~\bibnamefont {Gogol-Döring}}, \bibinfo {author} {\bibfnamefont {V.}~\bibnamefont {Kapitonov}}, \bibinfo {author} {\bibfnamefont {T.}~\bibnamefont {Diem}}, \bibinfo {author} {\bibfnamefont {A.}~\bibnamefont {Dalda}}, \bibinfo {author} {\bibfnamefont {J.}~\bibnamefont {Jurka}}, \bibinfo {author} {\bibfnamefont {E.}~\bibnamefont {Pritham}}, \bibinfo {author} {\bibfnamefont {F.}~\bibnamefont {Dyda}}, \bibinfo {author} {\bibfnamefont {Z.}~\bibnamefont {Izsvák}},\ and\ \bibinfo {author} {\bibfnamefont {Z.}~\bibnamefont {Ivics}},\ }\bibfield  {title} {\bibinfo {title} {A helitron
  transposon reconstructed from bats reveals a novel mechanism of genome shuffling in eukaryotes},\ }\href {https://doi.org/10.1038/ncomms10716} {\bibfield  {journal} {\bibinfo  {journal} {Nat Commun}\ }\textbf {\bibinfo {volume} {7}},\ \bibinfo {pages} {10716} (\bibinfo {year} {2016})},\ \bibinfo {note} {pMID: 26931494}\BibitemShut {NoStop}%
\bibitem [{\citenamefont {Grabundzija}\ \emph {et~al.}(2018)\citenamefont {Grabundzija}, \citenamefont {Hickman},\ and\ \citenamefont {Dyda}}]{Grabundzija_18_Dyda}%
  \BibitemOpen
  \bibfield  {author} {\bibinfo {author} {\bibfnamefont {I.}~\bibnamefont {Grabundzija}}, \bibinfo {author} {\bibfnamefont {A.}~\bibnamefont {Hickman}},\ and\ \bibinfo {author} {\bibfnamefont {F.}~\bibnamefont {Dyda}},\ }\bibfield  {title} {\bibinfo {title} {Helraiser intermediates provide insight into the mechanism of eukaryotic replicative transposition},\ }\href {https://doi.org/10.1038/s41467-018-03688-w} {\bibfield  {journal} {\bibinfo  {journal} {Nat Commun}\ }\textbf {\bibinfo {volume} {9(1)}},\ \bibinfo {pages} {1278} (\bibinfo {year} {2018})},\ \bibinfo {note} {pMID: 29599430}\BibitemShut {NoStop}%
\bibitem [{\citenamefont {Chakrabarty}\ \emph {et~al.}(2023)\citenamefont {Chakrabarty}, \citenamefont {Sen},\ and\ \citenamefont {Sengupta}}]{Chakrabarty_23_Sengupta}%
  \BibitemOpen
  \bibfield  {author} {\bibinfo {author} {\bibfnamefont {P.}~\bibnamefont {Chakrabarty}}, \bibinfo {author} {\bibfnamefont {R.}~\bibnamefont {Sen}},\ and\ \bibinfo {author} {\bibfnamefont {S.}~\bibnamefont {Sengupta}},\ }\bibfield  {title} {\bibinfo {title} {From parasites to partners: exploring the intricacies of host-transposon dynamics and coevolution},\ }\href {https://doi.org/10.1007/s10142-023-01206-w} {\bibfield  {journal} {\bibinfo  {journal} {Funct Integr Genomics}\ }\textbf {\bibinfo {volume} {23(3)}},\ \bibinfo {pages} {278} (\bibinfo {year} {2023})},\ \bibinfo {note} {pMID: 37610667}\BibitemShut {NoStop}%
\bibitem [{\citenamefont {Shen}\ \emph {et~al.}(1987)\citenamefont {Shen}, \citenamefont {Raleigh},\ and\ \citenamefont {Kleckner}}]{Shen_87_Kleckner}%
  \BibitemOpen
  \bibfield  {author} {\bibinfo {author} {\bibfnamefont {M.}~\bibnamefont {Shen}}, \bibinfo {author} {\bibfnamefont {E.}~\bibnamefont {Raleigh}},\ and\ \bibinfo {author} {\bibfnamefont {N.}~\bibnamefont {Kleckner}},\ }\bibfield  {title} {\bibinfo {title} {Physical analysis of tn10- and is10-promoted transpositions and rearrangements},\ }\href {https://doi.org/10.1093/genetics/116.3.359} {\bibfield  {journal} {\bibinfo  {journal} {Genetics}\ }\textbf {\bibinfo {volume} {116(3)}},\ \bibinfo {pages} {359} (\bibinfo {year} {1987})},\ \bibinfo {note} {pMID: 3038673}\BibitemShut {NoStop}%
\bibitem [{\citenamefont {Harada}\ \emph {et~al.}(1990)\citenamefont {Harada}, \citenamefont {Yukuhiro},\ and\ \citenamefont {Mukai}}]{Harada_90_Mukai}%
  \BibitemOpen
  \bibfield  {author} {\bibinfo {author} {\bibfnamefont {K.}~\bibnamefont {Harada}}, \bibinfo {author} {\bibfnamefont {K.}~\bibnamefont {Yukuhiro}},\ and\ \bibinfo {author} {\bibfnamefont {T.}~\bibnamefont {Mukai}},\ }\bibfield  {title} {\bibinfo {title} {Transposition rates of movable genetic elements in drosophila melanogaster},\ }\href {https://doi.org/10.1073/pnas.87.8.3248} {\bibfield  {journal} {\bibinfo  {journal} {Proc Natl Acad Sci U S A}\ }\textbf {\bibinfo {volume} {87(8)}},\ \bibinfo {pages} {3248} (\bibinfo {year} {1990})},\ \bibinfo {note} {pMID: 2158108}\BibitemShut {NoStop}%
\bibitem [{\citenamefont {Nuzhdin}\ and\ \citenamefont {Mackay}(1994)}]{Nuzhdin_94_Mackay}%
  \BibitemOpen
  \bibfield  {author} {\bibinfo {author} {\bibfnamefont {S.}~\bibnamefont {Nuzhdin}}\ and\ \bibinfo {author} {\bibfnamefont {T.}~\bibnamefont {Mackay}},\ }\bibfield  {title} {\bibinfo {title} {Direct determination of retrotransposon transposition rates in drosophila melanogaster},\ }\href {https://doi.org/10.1017/s0016672300032249} {\bibfield  {journal} {\bibinfo  {journal} {Genet Res}\ }\textbf {\bibinfo {volume} {63(2)}},\ \bibinfo {pages} {139} (\bibinfo {year} {1994})},\ \bibinfo {note} {pMID: 8026740}\BibitemShut {NoStop}%
\bibitem [{\citenamefont {Suh}\ \emph {et~al.}(1995)\citenamefont {Suh}, \citenamefont {Choi}, \citenamefont {Yamazaki},\ and\ \citenamefont {Harada}}]{Suh_95_Harada}%
  \BibitemOpen
  \bibfield  {author} {\bibinfo {author} {\bibfnamefont {D.}~\bibnamefont {Suh}}, \bibinfo {author} {\bibfnamefont {E.}~\bibnamefont {Choi}}, \bibinfo {author} {\bibfnamefont {T.}~\bibnamefont {Yamazaki}},\ and\ \bibinfo {author} {\bibfnamefont {K.}~\bibnamefont {Harada}},\ }\bibfield  {title} {\bibinfo {title} {Studies on the transposition rates of mobile genetic elements in a natural population of drosophila melanogaster},\ }\href {https://doi.org/10.1093/oxfordjournals.molbev.a040253} {\bibfield  {journal} {\bibinfo  {journal} {Mol Biol Evol}\ }\textbf {\bibinfo {volume} {12(5)}},\ \bibinfo {pages} {748} (\bibinfo {year} {1995})},\ \bibinfo {note} {pMID: 7476122}\BibitemShut {NoStop}%
\bibitem [{\citenamefont {Maside}\ \emph {et~al.}(2000)\citenamefont {Maside}, \citenamefont {Assimacopoulos},\ and\ \citenamefont {Charlesworth}}]{Maside_00_Charlesworth}%
  \BibitemOpen
  \bibfield  {author} {\bibinfo {author} {\bibfnamefont {X.}~\bibnamefont {Maside}}, \bibinfo {author} {\bibfnamefont {S.}~\bibnamefont {Assimacopoulos}},\ and\ \bibinfo {author} {\bibfnamefont {B.}~\bibnamefont {Charlesworth}},\ }\bibfield  {title} {\bibinfo {title} {Rates of movement of transposable elements on the second chromosome of drosophila melanogaster},\ }\href {https://doi.org/10.1017/s0016672399004474} {\bibfield  {journal} {\bibinfo  {journal} {Genet Res}\ }\textbf {\bibinfo {volume} {75(3)}},\ \bibinfo {pages} {275} (\bibinfo {year} {2000})},\ \bibinfo {note} {pMID: 10893864}\BibitemShut {NoStop}%
\bibitem [{\citenamefont {Sousa}\ \emph {et~al.}(2013)\citenamefont {Sousa}, \citenamefont {Bourgard}, \citenamefont {Wahl},\ and\ \citenamefont {Gordo}}]{Sousa_13_Gordo}%
  \BibitemOpen
  \bibfield  {author} {\bibinfo {author} {\bibfnamefont {A.}~\bibnamefont {Sousa}}, \bibinfo {author} {\bibfnamefont {C.}~\bibnamefont {Bourgard}}, \bibinfo {author} {\bibfnamefont {L.}~\bibnamefont {Wahl}},\ and\ \bibinfo {author} {\bibfnamefont {I.}~\bibnamefont {Gordo}},\ }\bibfield  {title} {\bibinfo {title} {Rates of transposition in escherichia coli},\ }\href {https://doi.org/10.1098/rsbl.2013.0838} {\bibfield  {journal} {\bibinfo  {journal} {Biol Lett}\ }\textbf {\bibinfo {volume} {9(6)}},\ \bibinfo {pages} {20130838} (\bibinfo {year} {2013})},\ \bibinfo {note} {pMID: 24307531}\BibitemShut {NoStop}%
\bibitem [{\citenamefont {Eanes}\ \emph {et~al.}(1988)\citenamefont {Eanes}, \citenamefont {Wesley}, \citenamefont {Hey}, \citenamefont {Houle},\ and\ \citenamefont {Ajioka}}]{Eanes_88_Ajioka}%
  \BibitemOpen
  \bibfield  {author} {\bibinfo {author} {\bibfnamefont {W.~F.}\ \bibnamefont {Eanes}}, \bibinfo {author} {\bibfnamefont {C.}~\bibnamefont {Wesley}}, \bibinfo {author} {\bibfnamefont {J.}~\bibnamefont {Hey}}, \bibinfo {author} {\bibfnamefont {D.}~\bibnamefont {Houle}},\ and\ \bibinfo {author} {\bibfnamefont {J.~W.}\ \bibnamefont {Ajioka}},\ }\bibfield  {title} {\bibinfo {title} {The fitness consequences of p element insertion in drosophila melanogaster},\ }\href {https://doi.org/10.1017/S0016672300027269} {\bibfield  {journal} {\bibinfo  {journal} {Genetics Research}\ }\textbf {\bibinfo {volume} {52}},\ \bibinfo {pages} {17–26} (\bibinfo {year} {1988})}\BibitemShut {NoStop}%
\bibitem [{\citenamefont {Charlesworth}\ and\ \citenamefont {Langley}(1989)}]{Charlesworth_89_Langley}%
  \BibitemOpen
  \bibfield  {author} {\bibinfo {author} {\bibfnamefont {B.}~\bibnamefont {Charlesworth}}\ and\ \bibinfo {author} {\bibfnamefont {C.}~\bibnamefont {Langley}},\ }\bibfield  {title} {\bibinfo {title} {The population genetics of drosophila transposable elements},\ }\href {https://doi.org/10.1146/annurev.ge.23.120189.001343} {\bibfield  {journal} {\bibinfo  {journal} {Annu Rev Genet}\ }\textbf {\bibinfo {volume} {23}},\ \bibinfo {pages} {251} (\bibinfo {year} {1989})},\ \bibinfo {note} {pMID: 2559652}\BibitemShut {NoStop}%
\bibitem [{\citenamefont {Houle}\ and\ \citenamefont {Nuzhdin}(2004)}]{Houle_04_Nuzhdin}%
  \BibitemOpen
  \bibfield  {author} {\bibinfo {author} {\bibfnamefont {D.}~\bibnamefont {Houle}}\ and\ \bibinfo {author} {\bibfnamefont {S.~V.}\ \bibnamefont {Nuzhdin}},\ }\bibfield  {title} {\bibinfo {title} {Mutation accumulation and the effect of copia insertions in drosophila melanogaster},\ }\href@noop {} {\bibfield  {journal} {\bibinfo  {journal} {Genetics Research}\ }\textbf {\bibinfo {volume} {83}},\ \bibinfo {pages} {7} (\bibinfo {year} {2004})}\BibitemShut {NoStop}%
\bibitem [{\citenamefont {Pasyukova}\ \emph {et~al.}(2004)\citenamefont {Pasyukova}, \citenamefont {Nuzhdin}, \citenamefont {Morozova},\ and\ \citenamefont {Mackay}}]{Pasyukova_04_Mackay}%
  \BibitemOpen
  \bibfield  {author} {\bibinfo {author} {\bibfnamefont {E.}~\bibnamefont {Pasyukova}}, \bibinfo {author} {\bibfnamefont {S.}~\bibnamefont {Nuzhdin}}, \bibinfo {author} {\bibfnamefont {T.}~\bibnamefont {Morozova}},\ and\ \bibinfo {author} {\bibfnamefont {T.}~\bibnamefont {Mackay}},\ }\bibfield  {title} {\bibinfo {title} {Accumulation of transposable elements in the genome of drosophila melanogaster is associated with a decrease in fitness},\ }\href@noop {} {\bibfield  {journal} {\bibinfo  {journal} {Journal of Heredity}\ }\textbf {\bibinfo {volume} {95}},\ \bibinfo {pages} {284} (\bibinfo {year} {2004})}\BibitemShut {NoStop}%
\bibitem [{\citenamefont {Steiniger}\ \emph {et~al.}(2006)\citenamefont {Steiniger}, \citenamefont {Adams}, \citenamefont {Marko},\ and\ \citenamefont {Reznikoff}}]{Steiniger_06_Reznikoff}%
  \BibitemOpen
  \bibfield  {author} {\bibinfo {author} {\bibfnamefont {M.}~\bibnamefont {Steiniger}}, \bibinfo {author} {\bibfnamefont {C.~D.}\ \bibnamefont {Adams}}, \bibinfo {author} {\bibfnamefont {J.~F.}\ \bibnamefont {Marko}},\ and\ \bibinfo {author} {\bibfnamefont {W.~S.}\ \bibnamefont {Reznikoff}},\ }\bibfield  {title} {\bibinfo {title} {Defining characteristics of tn 5 transposase non-specific dna binding},\ }\href {https://doi.org/10.1093/nar/gkl179} {\bibfield  {journal} {\bibinfo  {journal} {Nucleic Acids Research}\ }\textbf {\bibinfo {volume} {34}},\ \bibinfo {pages} {2820} (\bibinfo {year} {2006})}\BibitemShut {NoStop}%
\bibitem [{\citenamefont {Bire}\ \emph {et~al.}(2013)\citenamefont {Bire}, \citenamefont {Ley}, \citenamefont {Casteret}, \citenamefont {Mermod}, \citenamefont {Bigot},\ and\ \citenamefont {Rouleux-Bonnin}}]{Bire_13_RouleuxBonnin}%
  \BibitemOpen
  \bibfield  {author} {\bibinfo {author} {\bibfnamefont {S.}~\bibnamefont {Bire}}, \bibinfo {author} {\bibfnamefont {D.}~\bibnamefont {Ley}}, \bibinfo {author} {\bibfnamefont {S.}~\bibnamefont {Casteret}}, \bibinfo {author} {\bibfnamefont {N.}~\bibnamefont {Mermod}}, \bibinfo {author} {\bibfnamefont {Y.}~\bibnamefont {Bigot}},\ and\ \bibinfo {author} {\bibfnamefont {F.}~\bibnamefont {Rouleux-Bonnin}},\ }\bibfield  {title} {\bibinfo {title} {Optimization of the piggybac transposon using mrna and insulators: toward a more reliable gene delivery system},\ }\href {https://doi.org/10.1371/journal.pone.0082559} {\bibfield  {journal} {\bibinfo  {journal} {PLoS One}\ }\textbf {\bibinfo {volume} {8(12)}},\ \bibinfo {pages} {e82559} (\bibinfo {year} {2013})},\ \bibinfo {note} {pMID: 24312663}\BibitemShut {NoStop}%
\bibitem [{\citenamefont {Querques}\ \emph {et~al.}(2019)\citenamefont {Querques}, \citenamefont {Mades}, \citenamefont {Zuliani}, \citenamefont {Miskey}, \citenamefont {Alb}, \citenamefont {Grueso}, \citenamefont {Machwirth}, \citenamefont {Rausch}, \citenamefont {Einsele}, \citenamefont {Ivics}, \citenamefont {Hudecek},\ and\ \citenamefont {Barabas}}]{Querques_19_Barabas}%
  \BibitemOpen
  \bibfield  {author} {\bibinfo {author} {\bibfnamefont {I.}~\bibnamefont {Querques}}, \bibinfo {author} {\bibfnamefont {A.}~\bibnamefont {Mades}}, \bibinfo {author} {\bibfnamefont {C.}~\bibnamefont {Zuliani}}, \bibinfo {author} {\bibfnamefont {C.}~\bibnamefont {Miskey}}, \bibinfo {author} {\bibfnamefont {M.}~\bibnamefont {Alb}}, \bibinfo {author} {\bibfnamefont {E.}~\bibnamefont {Grueso}}, \bibinfo {author} {\bibfnamefont {M.}~\bibnamefont {Machwirth}}, \bibinfo {author} {\bibfnamefont {T.}~\bibnamefont {Rausch}}, \bibinfo {author} {\bibfnamefont {H.}~\bibnamefont {Einsele}}, \bibinfo {author} {\bibfnamefont {Z.}~\bibnamefont {Ivics}}, \bibinfo {author} {\bibfnamefont {M.}~\bibnamefont {Hudecek}},\ and\ \bibinfo {author} {\bibfnamefont {O.}~\bibnamefont {Barabas}},\ }\bibfield  {title} {\bibinfo {title} {A highly soluble sleeping beauty transposase improves control of gene insertion},\ }\href {https://doi.org/10.1038/s41587-019-0291-z} {\bibfield  {journal} {\bibinfo  {journal} {Nat Biotechnol}\ }\textbf
  {\bibinfo {volume} {37(12)}},\ \bibinfo {pages} {1502} (\bibinfo {year} {2019})},\ \bibinfo {note} {pMID: 31685959}\BibitemShut {NoStop}%
\bibitem [{\citenamefont {Yang}\ and\ \citenamefont {Bennetzen}(2009{\natexlab{a}})}]{Yang_09_Bennetzen}%
  \BibitemOpen
  \bibfield  {author} {\bibinfo {author} {\bibfnamefont {L.}~\bibnamefont {Yang}}\ and\ \bibinfo {author} {\bibfnamefont {J.~L.}\ \bibnamefont {Bennetzen}},\ }\bibfield  {title} {\bibinfo {title} {Structure-based discovery and description of plant and animal <i>helitrons</i>},\ }\href {https://doi.org/10.1073/pnas.0905563106} {\bibfield  {journal} {\bibinfo  {journal} {Proceedings of the National Academy of Sciences}\ }\textbf {\bibinfo {volume} {106}},\ \bibinfo {pages} {12832} (\bibinfo {year} {2009}{\natexlab{a}})},\ \Eprint {https://arxiv.org/abs/https://www.pnas.org/doi/pdf/10.1073/pnas.0905563106} {https://www.pnas.org/doi/pdf/10.1073/pnas.0905563106} \BibitemShut {NoStop}%
\bibitem [{\citenamefont {Yang}\ and\ \citenamefont {Bennetzen}(2009{\natexlab{b}})}]{Yang_09_Bennetzen_2}%
  \BibitemOpen
  \bibfield  {author} {\bibinfo {author} {\bibfnamefont {L.}~\bibnamefont {Yang}}\ and\ \bibinfo {author} {\bibfnamefont {J.~L.}\ \bibnamefont {Bennetzen}},\ }\bibfield  {title} {\bibinfo {title} {Distribution, diversity, evolution, and survival of <i>helitrons</i> in the maize genome},\ }\href {https://doi.org/10.1073/pnas.0908008106} {\bibfield  {journal} {\bibinfo  {journal} {Proceedings of the National Academy of Sciences}\ }\textbf {\bibinfo {volume} {106}},\ \bibinfo {pages} {19922} (\bibinfo {year} {2009}{\natexlab{b}})},\ \Eprint {https://arxiv.org/abs/https://www.pnas.org/doi/pdf/10.1073/pnas.0908008106} {https://www.pnas.org/doi/pdf/10.1073/pnas.0908008106} \BibitemShut {NoStop}%
\bibitem [{\citenamefont {Du}\ \emph {et~al.}(2009)\citenamefont {Du}, \citenamefont {Fefelova}, \citenamefont {Caronna}, \citenamefont {He},\ and\ \citenamefont {Dooner}}]{Du_09_Dooner}%
  \BibitemOpen
  \bibfield  {author} {\bibinfo {author} {\bibfnamefont {C.}~\bibnamefont {Du}}, \bibinfo {author} {\bibfnamefont {N.}~\bibnamefont {Fefelova}}, \bibinfo {author} {\bibfnamefont {J.}~\bibnamefont {Caronna}}, \bibinfo {author} {\bibfnamefont {L.}~\bibnamefont {He}},\ and\ \bibinfo {author} {\bibfnamefont {H.~K.}\ \bibnamefont {Dooner}},\ }\bibfield  {title} {\bibinfo {title} {The polychromatic <i>helitron</i> landscape of the maize genome},\ }\href {https://doi.org/10.1073/pnas.0904742106} {\bibfield  {journal} {\bibinfo  {journal} {Proceedings of the National Academy of Sciences}\ }\textbf {\bibinfo {volume} {106}},\ \bibinfo {pages} {19916} (\bibinfo {year} {2009})},\ \Eprint {https://arxiv.org/abs/https://www.pnas.org/doi/pdf/10.1073/pnas.0904742106} {https://www.pnas.org/doi/pdf/10.1073/pnas.0904742106} \BibitemShut {NoStop}%
\bibitem [{\citenamefont {Xiong}\ \emph {et~al.}(2014)\citenamefont {Xiong}, \citenamefont {He}, \citenamefont {Lai}, \citenamefont {Dooner},\ and\ \citenamefont {Du}}]{Xiong_14_Du}%
  \BibitemOpen
  \bibfield  {author} {\bibinfo {author} {\bibfnamefont {W.}~\bibnamefont {Xiong}}, \bibinfo {author} {\bibfnamefont {L.}~\bibnamefont {He}}, \bibinfo {author} {\bibfnamefont {J.}~\bibnamefont {Lai}}, \bibinfo {author} {\bibfnamefont {H.~K.}\ \bibnamefont {Dooner}},\ and\ \bibinfo {author} {\bibfnamefont {C.}~\bibnamefont {Du}},\ }\bibfield  {title} {\bibinfo {title} {Helitronscanner uncovers a large overlooked cache of <i>helitron</i> transposons in many plant genomes},\ }\href {https://doi.org/10.1073/pnas.1410068111} {\bibfield  {journal} {\bibinfo  {journal} {Proceedings of the National Academy of Sciences}\ }\textbf {\bibinfo {volume} {111}},\ \bibinfo {pages} {10263} (\bibinfo {year} {2014})},\ \Eprint {https://arxiv.org/abs/https://www.pnas.org/doi/pdf/10.1073/pnas.1410068111} {https://www.pnas.org/doi/pdf/10.1073/pnas.1410068111} \BibitemShut {NoStop}%
\bibitem [{\citenamefont {Cordaux}\ \emph {et~al.}(2006)\citenamefont {Cordaux}, \citenamefont {Hedges}, \citenamefont {Herke},\ and\ \citenamefont {Batzer}}]{Cordaux_06_Batzer}%
  \BibitemOpen
  \bibfield  {author} {\bibinfo {author} {\bibfnamefont {R.}~\bibnamefont {Cordaux}}, \bibinfo {author} {\bibfnamefont {D.~J.}\ \bibnamefont {Hedges}}, \bibinfo {author} {\bibfnamefont {S.~W.}\ \bibnamefont {Herke}},\ and\ \bibinfo {author} {\bibfnamefont {M.~A.}\ \bibnamefont {Batzer}},\ }\bibfield  {title} {\bibinfo {title} {Estimating the retrotransposition rate of human alu elements},\ }\href {https://doi.org/https://doi.org/10.1016/j.gene.2006.01.019} {\bibfield  {journal} {\bibinfo  {journal} {Gene}\ }\textbf {\bibinfo {volume} {373}},\ \bibinfo {pages} {134} (\bibinfo {year} {2006})}\BibitemShut {NoStop}%
\bibitem [{\citenamefont {Bennett}\ \emph {et~al.}(2008)\citenamefont {Bennett}, \citenamefont {Keller}, \citenamefont {Mills}, \citenamefont {Schmidt}, \citenamefont {Moran}, \citenamefont {Weichenrieder},\ and\ \citenamefont {Devine}}]{Bennett_08_Devine}%
  \BibitemOpen
  \bibfield  {author} {\bibinfo {author} {\bibfnamefont {E.}~\bibnamefont {Bennett}}, \bibinfo {author} {\bibfnamefont {H.}~\bibnamefont {Keller}}, \bibinfo {author} {\bibfnamefont {R.}~\bibnamefont {Mills}}, \bibinfo {author} {\bibfnamefont {S.}~\bibnamefont {Schmidt}}, \bibinfo {author} {\bibfnamefont {J.}~\bibnamefont {Moran}}, \bibinfo {author} {\bibfnamefont {O.}~\bibnamefont {Weichenrieder}},\ and\ \bibinfo {author} {\bibfnamefont {S.}~\bibnamefont {Devine}},\ }\bibfield  {title} {\bibinfo {title} {Active alu retrotransposons in the human genome},\ }\href {https://doi.org/10.1101/gr.081737.108} {\bibfield  {journal} {\bibinfo  {journal} {Genome Res}\ }\textbf {\bibinfo {volume} {18(12)}},\ \bibinfo {pages} {1875} (\bibinfo {year} {2008})},\ \bibinfo {note} {pMID: 18836035}\BibitemShut {NoStop}%
\bibitem [{\citenamefont {Xing}\ \emph {et~al.}(2009)\citenamefont {Xing}, \citenamefont {Zhang}, \citenamefont {Han}, \citenamefont {Salem}, \citenamefont {Sen}, \citenamefont {Huff}, \citenamefont {Zhou}, \citenamefont {Kirkness}, \citenamefont {Levy}, \citenamefont {Batzer},\ and\ \citenamefont {Jorde}}]{Xing_09_Jorde}%
  \BibitemOpen
  \bibfield  {author} {\bibinfo {author} {\bibfnamefont {J.}~\bibnamefont {Xing}}, \bibinfo {author} {\bibfnamefont {Y.}~\bibnamefont {Zhang}}, \bibinfo {author} {\bibfnamefont {K.}~\bibnamefont {Han}}, \bibinfo {author} {\bibfnamefont {A.}~\bibnamefont {Salem}}, \bibinfo {author} {\bibfnamefont {S.}~\bibnamefont {Sen}}, \bibinfo {author} {\bibfnamefont {C.}~\bibnamefont {Huff}}, \bibinfo {author} {\bibfnamefont {Q.}~\bibnamefont {Zhou}}, \bibinfo {author} {\bibfnamefont {E.}~\bibnamefont {Kirkness}}, \bibinfo {author} {\bibfnamefont {S.}~\bibnamefont {Levy}}, \bibinfo {author} {\bibfnamefont {M.}~\bibnamefont {Batzer}},\ and\ \bibinfo {author} {\bibfnamefont {L.}~\bibnamefont {Jorde}},\ }\bibfield  {title} {\bibinfo {title} {Mobile elements create structural variation: analysis of a complete human genome},\ }\href {https://doi.org/10.1101/gr.091827.109} {\bibfield  {journal} {\bibinfo  {journal} {Genome Res}\ }\textbf {\bibinfo {volume} {19(9)}},\ \bibinfo {pages} {1516} (\bibinfo {year} {2009})},\ \bibinfo
  {note} {pMID: 19439515}\BibitemShut {NoStop}%
\bibitem [{\citenamefont {Beck}\ \emph {et~al.}(2011)\citenamefont {Beck}, \citenamefont {Garcia-Perez}, \citenamefont {Badge},\ and\ \citenamefont {Moran}}]{Beck_11_Moran}%
  \BibitemOpen
  \bibfield  {author} {\bibinfo {author} {\bibfnamefont {C.}~\bibnamefont {Beck}}, \bibinfo {author} {\bibfnamefont {J.}~\bibnamefont {Garcia-Perez}}, \bibinfo {author} {\bibfnamefont {R.}~\bibnamefont {Badge}},\ and\ \bibinfo {author} {\bibfnamefont {J.}~\bibnamefont {Moran}},\ }\bibfield  {title} {\bibinfo {title} {Line-1 elements in structural variation and disease},\ }\href {https://doi.org/10.1146/annurev-genom-082509-141802} {\bibfield  {journal} {\bibinfo  {journal} {Annu Rev Genomics Hum Genet}\ }\textbf {\bibinfo {volume} {12}},\ \bibinfo {pages} {187} (\bibinfo {year} {2011})},\ \bibinfo {note} {pMID: 21801021}\BibitemShut {NoStop}%
\bibitem [{\citenamefont {Esnault}\ \emph {et~al.}(2000)\citenamefont {Esnault}, \citenamefont {Maestre},\ and\ \citenamefont {Heidmann}}]{Esnault_00_Heidmann}%
  \BibitemOpen
  \bibfield  {author} {\bibinfo {author} {\bibfnamefont {C.}~\bibnamefont {Esnault}}, \bibinfo {author} {\bibfnamefont {J.}~\bibnamefont {Maestre}},\ and\ \bibinfo {author} {\bibfnamefont {T.}~\bibnamefont {Heidmann}},\ }\bibfield  {title} {\bibinfo {title} {Human line retrotransposons generate processed pseudogenes},\ }\href {https://doi.org/10.1038/74184} {\bibfield  {journal} {\bibinfo  {journal} {Nat Genet}\ }\textbf {\bibinfo {volume} {24(4)}},\ \bibinfo {pages} {363} (\bibinfo {year} {2000})},\ \bibinfo {note} {pMID: 10742098}\BibitemShut {NoStop}%
\bibitem [{\citenamefont {Wei}\ \emph {et~al.}(2001)\citenamefont {Wei}, \citenamefont {Gilbert}, \citenamefont {Ooi}, \citenamefont {Lawler}, \citenamefont {Ostertag}, \citenamefont {Kazazian}, \citenamefont {Boeke},\ and\ \citenamefont {Moran}}]{Wei_01_Moran}%
  \BibitemOpen
  \bibfield  {author} {\bibinfo {author} {\bibfnamefont {W.}~\bibnamefont {Wei}}, \bibinfo {author} {\bibfnamefont {N.}~\bibnamefont {Gilbert}}, \bibinfo {author} {\bibfnamefont {S.~L.}\ \bibnamefont {Ooi}}, \bibinfo {author} {\bibfnamefont {J.~F.}\ \bibnamefont {Lawler}}, \bibinfo {author} {\bibfnamefont {E.~M.}\ \bibnamefont {Ostertag}}, \bibinfo {author} {\bibfnamefont {H.~H.}\ \bibnamefont {Kazazian}}, \bibinfo {author} {\bibfnamefont {J.~D.}\ \bibnamefont {Boeke}},\ and\ \bibinfo {author} {\bibfnamefont {J.~V.}\ \bibnamefont {Moran}},\ }\bibfield  {title} {\bibinfo {title} {Human l1 retrotransposition: cis preference versus trans complementation},\ }\href {https://doi.org/10.1128/MCB.21.4.1429-1439.2001} {\bibfield  {journal} {\bibinfo  {journal} {Molecular and Cellular Biology}\ }\textbf {\bibinfo {volume} {21}},\ \bibinfo {pages} {1429} (\bibinfo {year} {2001})},\ \bibinfo {note} {pMID: 11158327},\ \Eprint {https://arxiv.org/abs/https://doi.org/10.1128/MCB.21.4.1429-1439.2001}
  {https://doi.org/10.1128/MCB.21.4.1429-1439.2001} \BibitemShut {NoStop}%
\bibitem [{\citenamefont {Bolton}\ \emph {et~al.}(2005)\citenamefont {Bolton}, \citenamefont {Coombes}, \citenamefont {Eby}, \citenamefont {Cardell},\ and\ \citenamefont {Boeke}}]{Bolton_05_Boeke}%
  \BibitemOpen
  \bibfield  {author} {\bibinfo {author} {\bibfnamefont {E.}~\bibnamefont {Bolton}}, \bibinfo {author} {\bibfnamefont {C.}~\bibnamefont {Coombes}}, \bibinfo {author} {\bibfnamefont {Y.}~\bibnamefont {Eby}}, \bibinfo {author} {\bibfnamefont {M.}~\bibnamefont {Cardell}},\ and\ \bibinfo {author} {\bibfnamefont {J.}~\bibnamefont {Boeke}},\ }\bibfield  {title} {\bibinfo {title} {Identification and characterization of critical cis-acting sequences within the yeast ty1 retrotransposon},\ }\href {https://doi.org/10.1261/rna.7860605} {\bibfield  {journal} {\bibinfo  {journal} {RNA}\ }\textbf {\bibinfo {volume} {11(3)}},\ \bibinfo {pages} {308} (\bibinfo {year} {2005})},\ \bibinfo {note} {pMID: 15661848}\BibitemShut {NoStop}%
\bibitem [{\citenamefont {Curcio}\ \emph {et~al.}(2015)\citenamefont {Curcio}, \citenamefont {Lutz},\ and\ \citenamefont {Lesage}}]{Curcio_15_Lesage}%
  \BibitemOpen
  \bibfield  {author} {\bibinfo {author} {\bibfnamefont {M.~J.}\ \bibnamefont {Curcio}}, \bibinfo {author} {\bibfnamefont {S.}~\bibnamefont {Lutz}},\ and\ \bibinfo {author} {\bibfnamefont {P.}~\bibnamefont {Lesage}},\ }\bibfield  {title} {\bibinfo {title} {The ty1 ltr-retrotransposon of budding yeast, <i>saccharomyces cerevisiae</i>},\ }\href {https://doi.org/10.1128/microbiolspec.mdna3-0053-2014} {\bibfield  {journal} {\bibinfo  {journal} {Microbiology Spectrum}\ }\textbf {\bibinfo {volume} {3}},\ \bibinfo {pages} {10.1128/microbiolspec.mdna3} (\bibinfo {year} {2015})},\ \Eprint {https://arxiv.org/abs/https://journals.asm.org/doi/pdf/10.1128/microbiolspec.mdna3-0053-2014} {https://journals.asm.org/doi/pdf/10.1128/microbiolspec.mdna3-0053-2014} \BibitemShut {NoStop}%
\bibitem [{\citenamefont {GREGORY}(2005)}]{Gregory_05}%
  \BibitemOpen
  \bibfield  {author} {\bibinfo {author} {\bibfnamefont {T.~R.}\ \bibnamefont {GREGORY}},\ }\bibfield  {title} {\bibinfo {title} {The c-value enigma in plants and animals: A review of parallels and an appeal for partnership},\ }\href {https://doi.org/10.1093/aob/mci009} {\bibfield  {journal} {\bibinfo  {journal} {Annals of Botany}\ }\textbf {\bibinfo {volume} {95}},\ \bibinfo {pages} {133} (\bibinfo {year} {2005})}\BibitemShut {NoStop}%
\bibitem [{\citenamefont {Brookfield}(1982)}]{Brookfield_82}%
  \BibitemOpen
  \bibfield  {author} {\bibinfo {author} {\bibfnamefont {J.}~\bibnamefont {Brookfield}},\ }\bibfield  {title} {\bibinfo {title} {Interspersed repetitive dna sequences are unlikely to be parasitic},\ }\href {https://doi.org/10.1016/0022-5193(82)90313-7} {\bibfield  {journal} {\bibinfo  {journal} {J Theor Biol}\ }\textbf {\bibinfo {volume} {94(2)}},\ \bibinfo {pages} {281} (\bibinfo {year} {1982})},\ \bibinfo {note} {pMID: 7078209}\BibitemShut {NoStop}%
\bibitem [{\citenamefont {Charlesworth}\ and\ \citenamefont {Charlesworth}(1983)}]{Charlesworth_83_Charlesworth}%
  \BibitemOpen
  \bibfield  {author} {\bibinfo {author} {\bibfnamefont {B.}~\bibnamefont {Charlesworth}}\ and\ \bibinfo {author} {\bibfnamefont {D.}~\bibnamefont {Charlesworth}},\ }\bibfield  {title} {\bibinfo {title} {The population dynamics of transposable elements},\ }\href {https://doi.org/10.1017/S0016672300021455} {\bibfield  {journal} {\bibinfo  {journal} {Genetics Research}\ }\textbf {\bibinfo {volume} {42(1)}},\ \bibinfo {pages} {1–27} (\bibinfo {year} {1983})}\BibitemShut {NoStop}%
\bibitem [{\citenamefont {Montgomery}\ \emph {et~al.}(1987)\citenamefont {Montgomery}, \citenamefont {Charlesworth},\ and\ \citenamefont {Langley}}]{Montgomery_87_Langley}%
  \BibitemOpen
  \bibfield  {author} {\bibinfo {author} {\bibfnamefont {E.}~\bibnamefont {Montgomery}}, \bibinfo {author} {\bibfnamefont {B.}~\bibnamefont {Charlesworth}},\ and\ \bibinfo {author} {\bibfnamefont {C.}~\bibnamefont {Langley}},\ }\bibfield  {title} {\bibinfo {title} {A test for the role of natural selection in the stabilization of transposable element copy number in a population of drosophila melanogaster},\ }\href {https://doi.org/10.1017/s0016672300026707} {\bibfield  {journal} {\bibinfo  {journal} {Genet Res}\ }\textbf {\bibinfo {volume} {49(1)}},\ \bibinfo {pages} {31} (\bibinfo {year} {1987})},\ \bibinfo {note} {pMID: 3032743}\BibitemShut {NoStop}%
\bibitem [{\citenamefont {Langley}\ \emph {et~al.}(1988)\citenamefont {Langley}, \citenamefont {Montgomery}, \citenamefont {Hudson}, \citenamefont {Kaplan},\ and\ \citenamefont {Charlesworth}}]{Langley_88_Charlesworth}%
  \BibitemOpen
  \bibfield  {author} {\bibinfo {author} {\bibfnamefont {C.}~\bibnamefont {Langley}}, \bibinfo {author} {\bibfnamefont {E.}~\bibnamefont {Montgomery}}, \bibinfo {author} {\bibfnamefont {R.}~\bibnamefont {Hudson}}, \bibinfo {author} {\bibfnamefont {N.}~\bibnamefont {Kaplan}},\ and\ \bibinfo {author} {\bibfnamefont {B.}~\bibnamefont {Charlesworth}},\ }\bibfield  {title} {\bibinfo {title} {On the role of unequal exchange in the containment of transposable element copy number},\ }\href {https://doi.org/10.1017/s0016672300027695} {\bibfield  {journal} {\bibinfo  {journal} {Genet Res}\ }\textbf {\bibinfo {volume} {52(3)}},\ \bibinfo {pages} {223} (\bibinfo {year} {1988})},\ \bibinfo {note} {pMID: 2854088}\BibitemShut {NoStop}%
\bibitem [{\citenamefont {Delprat}\ \emph {et~al.}(2009)\citenamefont {Delprat}, \citenamefont {Negre}, \citenamefont {Puig},\ and\ \citenamefont {Ruiz}}]{Delprat_09_Ruiz}%
  \BibitemOpen
  \bibfield  {author} {\bibinfo {author} {\bibfnamefont {A.}~\bibnamefont {Delprat}}, \bibinfo {author} {\bibfnamefont {B.}~\bibnamefont {Negre}}, \bibinfo {author} {\bibfnamefont {M.}~\bibnamefont {Puig}},\ and\ \bibinfo {author} {\bibfnamefont {A.}~\bibnamefont {Ruiz}},\ }\bibfield  {title} {\bibinfo {title} {The transposon galileo generates natural chromosomal inversions in drosophila by ectopic recombination},\ }\href {https://doi.org/10.1371/journal.pone.0007883} {\bibfield  {journal} {\bibinfo  {journal} {PLoS One}\ }\textbf {\bibinfo {volume} {4(11)}},\ \bibinfo {pages} {e7883} (\bibinfo {year} {2009})},\ \bibinfo {note} {pMID: 19936241}\BibitemShut {NoStop}%
\bibitem [{\citenamefont {Santana}\ \emph {et~al.}(2014)\citenamefont {Santana}, \citenamefont {Silva}, \citenamefont {Mizubuti}, \citenamefont {Araújo}, \citenamefont {Condon}, \citenamefont {Turgeon},\ and\ \citenamefont {Queiroz}}]{Santana_14_Queiroz}%
  \BibitemOpen
  \bibfield  {author} {\bibinfo {author} {\bibfnamefont {M.}~\bibnamefont {Santana}}, \bibinfo {author} {\bibfnamefont {J.}~\bibnamefont {Silva}}, \bibinfo {author} {\bibfnamefont {E.}~\bibnamefont {Mizubuti}}, \bibinfo {author} {\bibfnamefont {E.}~\bibnamefont {Araújo}}, \bibinfo {author} {\bibfnamefont {B.}~\bibnamefont {Condon}}, \bibinfo {author} {\bibfnamefont {B.}~\bibnamefont {Turgeon}},\ and\ \bibinfo {author} {\bibfnamefont {M.}~\bibnamefont {Queiroz}},\ }\bibfield  {title} {\bibinfo {title} {Characterization and potential evolutionary impact of transposable elements in the genome of cochliobolus heterostrophus},\ }\href {https://doi.org/10.1186/1471-2164-15-536} {\bibfield  {journal} {\bibinfo  {journal} {BMC Genomics}\ }\textbf {\bibinfo {volume} {15(1)}},\ \bibinfo {pages} {536} (\bibinfo {year} {2014})},\ \bibinfo {note} {pMID: 24973942}\BibitemShut {NoStop}%
\bibitem [{\citenamefont {Voliva}\ \emph {et~al.}(1983)\citenamefont {Voliva}, \citenamefont {Jahn}, \citenamefont {Comer}, \citenamefont {Hutchison},\ and\ \citenamefont {Edgell}}]{Voliva_83_Edgell}%
  \BibitemOpen
  \bibfield  {author} {\bibinfo {author} {\bibfnamefont {C.~F.}\ \bibnamefont {Voliva}}, \bibinfo {author} {\bibfnamefont {C.~L.}\ \bibnamefont {Jahn}}, \bibinfo {author} {\bibfnamefont {M.~B.}\ \bibnamefont {Comer}}, \bibinfo {author} {\bibfnamefont {I.}~\bibnamefont {Hutchison}, \bibfnamefont {Clyde~A.}},\ and\ \bibinfo {author} {\bibfnamefont {M.~H.}\ \bibnamefont {Edgell}},\ }\bibfield  {title} {\bibinfo {title} {{The L1Md long interspersed repeat family in the mouse: almost all examples are truncated at one end}},\ }\href {https://doi.org/10.1093/nar/11.24.8847} {\bibfield  {journal} {\bibinfo  {journal} {Nucleic Acids Research}\ }\textbf {\bibinfo {volume} {11}},\ \bibinfo {pages} {8847} (\bibinfo {year} {1983})},\ \Eprint {https://arxiv.org/abs/https://academic.oup.com/nar/article-pdf/11/24/8847/7048814/11-24-8847.pdf} {https://academic.oup.com/nar/article-pdf/11/24/8847/7048814/11-24-8847.pdf} \BibitemShut {NoStop}%
\bibitem [{\citenamefont {Hardies}\ \emph {et~al.}(1986)\citenamefont {Hardies}, \citenamefont {Martin}, \citenamefont {Voliva}, \citenamefont {Hutchison},\ and\ \citenamefont {Edgell}}]{Hardies_86_Edgell}%
  \BibitemOpen
  \bibfield  {author} {\bibinfo {author} {\bibfnamefont {S.~C.}\ \bibnamefont {Hardies}}, \bibinfo {author} {\bibfnamefont {S.~L.}\ \bibnamefont {Martin}}, \bibinfo {author} {\bibfnamefont {C.~F.}\ \bibnamefont {Voliva}}, \bibinfo {author} {\bibfnamefont {r.}~\bibnamefont {Hutchison}, \bibfnamefont {C~A}},\ and\ \bibinfo {author} {\bibfnamefont {M.~H.}\ \bibnamefont {Edgell}},\ }\bibfield  {title} {\bibinfo {title} {{An analysis of replacement and synonymous changes in the rodent L1 repeat family.}},\ }\href {https://doi.org/10.1093/oxfordjournals.molbev.a040386} {\bibfield  {journal} {\bibinfo  {journal} {Molecular Biology and Evolution}\ }\textbf {\bibinfo {volume} {3}},\ \bibinfo {pages} {109} (\bibinfo {year} {1986})},\ \Eprint {https://arxiv.org/abs/https://academic.oup.com/mbe/article-pdf/3/2/109/11167308/2hard.pdf} {https://academic.oup.com/mbe/article-pdf/3/2/109/11167308/2hard.pdf} \BibitemShut {NoStop}%
\bibitem [{\citenamefont {Sassaman}\ \emph {et~al.}(1997)\citenamefont {Sassaman}, \citenamefont {Dombroski}, \citenamefont {Moran}, \citenamefont {Kimberland}, \citenamefont {Naas}, \citenamefont {DeBerardinis}, \citenamefont {Gabriel}, \citenamefont {Swergold},\ and\ \citenamefont {Kazazian}}]{Sassaman_97_Kazazian}%
  \BibitemOpen
  \bibfield  {author} {\bibinfo {author} {\bibfnamefont {D.}~\bibnamefont {Sassaman}}, \bibinfo {author} {\bibfnamefont {B.}~\bibnamefont {Dombroski}}, \bibinfo {author} {\bibfnamefont {J.}~\bibnamefont {Moran}}, \bibinfo {author} {\bibfnamefont {M.}~\bibnamefont {Kimberland}}, \bibinfo {author} {\bibfnamefont {T.}~\bibnamefont {Naas}}, \bibinfo {author} {\bibfnamefont {R.}~\bibnamefont {DeBerardinis}}, \bibinfo {author} {\bibfnamefont {A.}~\bibnamefont {Gabriel}}, \bibinfo {author} {\bibfnamefont {G.}~\bibnamefont {Swergold}},\ and\ \bibinfo {author} {\bibfnamefont {H.~J.}\ \bibnamefont {Kazazian}},\ }\bibfield  {title} {\bibinfo {title} {Many human l1 elements are capable of retrotransposition},\ }\href {https://doi.org/10.1038/ng0597-37} {\bibfield  {journal} {\bibinfo  {journal} {Nat Genet}\ }\textbf {\bibinfo {volume} {16(1)}},\ \bibinfo {pages} {37} (\bibinfo {year} {1997})},\ \bibinfo {note} {pMID: 9140393}\BibitemShut {NoStop}%
\bibitem [{\citenamefont {Ostertag}\ and\ \citenamefont {Kazazian}(2001)}]{Ostertag_01_Kazazian}%
  \BibitemOpen
  \bibfield  {author} {\bibinfo {author} {\bibfnamefont {E.}~\bibnamefont {Ostertag}}\ and\ \bibinfo {author} {\bibfnamefont {H.~J.}\ \bibnamefont {Kazazian}},\ }\bibfield  {title} {\bibinfo {title} {Twin priming: a proposed mechanism for the creation of inversions in l1 retrotransposition},\ }\href {https://doi.org/10.1101/gr.205701} {\bibfield  {journal} {\bibinfo  {journal} {Genome Res}\ }\textbf {\bibinfo {volume} {11(12)}},\ \bibinfo {pages} {2059} (\bibinfo {year} {2001})},\ \bibinfo {note} {pMID: 11731496}\BibitemShut {NoStop}%
\bibitem [{\citenamefont {Brouha}\ \emph {et~al.}(2003)\citenamefont {Brouha}, \citenamefont {Schustak}, \citenamefont {Badge}, \citenamefont {Lutz-Prigge}, \citenamefont {Farley}, \citenamefont {Moran},\ and\ \citenamefont {Kazazian}}]{Brouha_03_Kazazian}%
  \BibitemOpen
  \bibfield  {author} {\bibinfo {author} {\bibfnamefont {B.}~\bibnamefont {Brouha}}, \bibinfo {author} {\bibfnamefont {J.}~\bibnamefont {Schustak}}, \bibinfo {author} {\bibfnamefont {R.}~\bibnamefont {Badge}}, \bibinfo {author} {\bibfnamefont {S.}~\bibnamefont {Lutz-Prigge}}, \bibinfo {author} {\bibfnamefont {A.}~\bibnamefont {Farley}}, \bibinfo {author} {\bibfnamefont {J.}~\bibnamefont {Moran}},\ and\ \bibinfo {author} {\bibfnamefont {H.~J.}\ \bibnamefont {Kazazian}},\ }\bibfield  {title} {\bibinfo {title} {Hot l1s account for the bulk of retrotransposition in the human population},\ }\href {https://doi.org/10.1073/pnas.0831042100} {\bibfield  {journal} {\bibinfo  {journal} {Proc Natl Acad Sci U S A}\ }\textbf {\bibinfo {volume} {100(9)}},\ \bibinfo {pages} {5280} (\bibinfo {year} {2003})},\ \bibinfo {note} {pMID: 12682288}\BibitemShut {NoStop}%
\end{thebibliography}%

\end{document}